\documentclass[12pt]{iopart}
\usepackage{iopams}  
\usepackage{graphicx,comment,amssymb}
\usepackage[labelfont=bf, labelsep=period]{caption}

\usepackage[square, numbers, sort&compress]{natbib}

\usepackage{hyperref}
\hypersetup{
    colorlinks=true,
    citecolor=blue,
    urlcolor=blue,
    linkcolor=black,
}

\begin{document}

\paper[Nonradiating anapole states in nanophotonics]{Nonradiating anapole states in nanophotonics: from fundamentals to applications}

\author{Yuanqing Yang \& Sergey I. Bozhevolnyi}

\address{Centre for Nano Optics, University of Southern Denmark, Campusvej 55, DK-5230 Odense M, Denmark}
\ead{\href{mailto:yy@mci.sdu.dk}{yy@mci.sdu.dk}}
\vspace{10pt}
\begin{indented}
\item[]September 2018
\end{indented}

\begin{abstract} Nonradiating sources are nontrivial charge-current distributions that do not generate fields outside the source domain. The pursuit of their possible existence has fascinated several generations of physicists and triggered developments in various branches of science ranging from medical imaging to dark matter. Recently, one of the most fundamental types of nonradiating sources, named anapole states, has been realized in nanophotonics regime and soon spurred considerable research efforts and widespread interest. A series of astounding advances have been achieved within a very short period of time, uncovering the great potential of anapole states in many aspects such as lasing, sensing, metamaterials, and nonlinear optics. In this review, we provide a detailed account of anapole states in nanophotonics research, encompassing their basic concepts, historical origins, and new physical effects. We discuss the recent research frontiers in understanding and employing optical anapoles and provide an outlook for this vibrant field of research.

\vspace{1pc}
\end{abstract}

%Uncomment for keywords

%Uncomment for Submitted to journal title message
% \maketitle

\section{Introduction}

Conventional wisdom, even to this day, states that accelerating charges should radiate electromagnetic energy \cite{jackson_classical_1999}. Indeed, this notion has stimulated the development of a diverse set of modern technologies, ranging from consumer electronics to large-scale particle accelerators. However, from the early days of electromagnetic theory, scientists have started to envision the possible existence of \emph{nonradiating sources}, i.e., extended charge-current distributions that do not produce radiation fields outside their domains \cite{ehrenfest1910ungleichformige}. To date, various types of nonradiating sources and relevant general theories have been proposed \cite{schott1933lix, bohm1948self, goedecke1964classically, devaney1973radiating, van1991singular, nikolova2005nonradiating, savinov2018light}. In this context, a special type of configurations, termed anapole states, has attracted the attention of many physicists as it may serve as an elementary building block of myriad nonradiating sources.

The concept of anapole states was first introduced by Zeldovich \cite{zel1958electromagnetic} in particle physics. In 1957, he pointed out that the weak interactions in elementary particles would lead to a new parity-violating moment, which he called an 'anapole' moment (means 'without poles' in Greek).  Since then, the anapole moment has made its impact on several branches of physics due to its intriguing properties yet a simple configuration. For instance, it could be used as a delicate probe to observe atomic parity nonconservation in nuclei \cite{wood1997measurement} and was also suggested as the only allowed electromagnetic moment in Majorana fermions to explain the elusive nature of dark matter \cite{radescu1985electromagnetic, ho2013anapole, gao2014anapole}. 

In electrodynamics, oscillating anapoles are usually understood by means of multipole expansion, in which time-varying charge-current distributions can be presented by a series of point-like multipole sources. The simplest case of an anapole could be seen as the superposition of an electric dipole and a toroidal dipole moment \cite{miroshnichenko2015nonradiating}. The identical radiation patterns of the two dipoles give rise to a completely destructive interference in the far field and the corresponding radiationless behavior of the anapole. Recently, advances in precise nanofabrication and metamaterials enabled the unveiling of such phenomena in a wide range of the electromagnetic spectrum, spanning from microwave to near-infrared and visible frequencies \cite{fedotov2013resonant, miroshnichenko2015nonradiating, gongora2017anapole, zenin2017direct}. In these well-designed structures, the existence of anapole states brings in not only suppression in far-field radiation but also enhancements in near-field energy, which may facilitate many physical processes and open new venues for a variety of relevant applications such as cloaking, sensing, metamaterials, and nonlinear optics. 

Here we review the state-of-art research on anapole states with a strong focus on the regime of nanophotonics. To provide a solid background, we will begin with a brief history of anapole moments and show how these concepts were transplanted from atomic physics into modern nanophotonics. Then we start out to examine the physical origin of anapoles within the framework of multipole expansion. What follows is an overview of recent developments of anapole states and their applications. Finally, we conclude our discussions and provide an outlook for the outstanding opportunities in this rapidly developing field.

\section{Background and Fundamentals}

\subsection{History: from nonradiating sources to static and dynamic anapoles}

Anapole states are usually referred to as a type of nonradiating sources. The history of seeking such nonradiating sources can be traced back to the early days after J. J. Thomson's discovery of electrons. Initiated by Sommerfeld, Herglotz, and Hertz \cite{herglotz1903elektronentheorie, sommerfeld1904, sommerfeld1904e, sommerfeld1905e, herglotz1907integralgleichungen, hertz1907bewegung}, several extended electron models had been considered to describe the atomic structure. However, all these classical models were facing one simple problem: the inclusion of electric charges in accelerated motion. Given the retarded potential solution to Maxwell equations, it was a common understanding in that era that accelerating charges necessarily radiate energy in the form of electromagnetic waves. As such, the atoms described by all these models would constantly release energy and collapse very quickly. The scientists thus started to visualize if there exist nontrivial charge distributions which do not radiate. In 1910, Paul Ehrenfest first explicitly demonstrated that there exist at least two extended charge distributions which can accelerate but not generate radiation \cite{ehrenfest1910ungleichformige}.

Whereas the subsequent proposal of the Bohr's model in 1913 soon made Ehrenfest's interesting work almost forgotten in time, there was still a small stream of scientists continuing along this line. In 1933, Schott \cite{schott1933lix} showed that it is possible for a charged spherical shell to move in a periodic orbit without producing any radiation. Later, Bohm and Weinstein \cite{bohm1948self} extended Schott’s treatment to other spherically symmetric distributions oscillating linearly. In 1964, Goedecke \cite{goedecke1964classically} developed a more general theory for nonradiating sources and exemplified his theory with a few asymmetric or spinning configurations. In more modern times, nonradiating sources are often linked to the inverse source problem \cite{bleistein1977nonuniqueness, hoenders1997existence, nikolova2005nonradiating}, that is, how to determine the internal distribution of a radiating source from its emitted far field. The general existence and various classes of nonradiating configurations imply the nonuniqueness of the solutions, which have fundamentally impacted on many research areas from medical imaging to astronomy and radar science \cite{devaney2012mathematical}.

The early demonstrations of nonradiating sources were often used to explain the fundamental structures or properties of elementary particles. But this was not limited within the scope of classical models. In quantum field theory, shortly after the discovery of parity violation, Zel’dovich \cite{zel1958electromagnetic} predicted that the effects of parity nonconservation on electric charges would result in a new kind of electromagnetic interaction and associated charge-current distributions, which he named as an 'anapole'. In contrast to the nuclear electric dipole moment which violates both parity (\textit{P}) and time-reversal (\textit{T}) invariance, the \emph{static} anapole moment violates \textit{P} but conserves \textit{T}. Zel'dovich further vividly described the anapole distribution as a wire solenoid curved around into a torus, where a static toroidal current creates a magnetic field only inside the toroid and an electric field is everywhere zero. However, despite its interesting physical manifestation, the anapole moment of elementary particles was too small to be observed under the experimental condition at that time. Therefore, it was only decades later when high precision atomic experiment seemed to be feasible, considerable interest had once again been aroused in the 'Zeldovich's anapole'.  Several theoretical works \cite{flambaum1980p, flambaum1984nuclear, haxton1989nucleon, bouchiat1991nuclear} showed that a static anapole moment could serve as a valuable probe to detect the atomic parity nonconservation effect in nuclei.  Finally, in 1997, 40 years after its first theoretical prediction, the long-sought static anapole moment, for the first time, was experimentally observed and measured by Wood \textit{et al}. in atomic Cesium \cite{wood1997measurement}. This inspiring result later stimulated a plethora of investigations on the static anapole moment in the context of nuclear and molecular physics \cite{flambaum1997anapole, ceulemans1998molecular, haxton2002nuclear, pelloni2011magnetic, safronova2018search}. 

\begin{figure}[h]
\centering
\includegraphics[width=8cm]{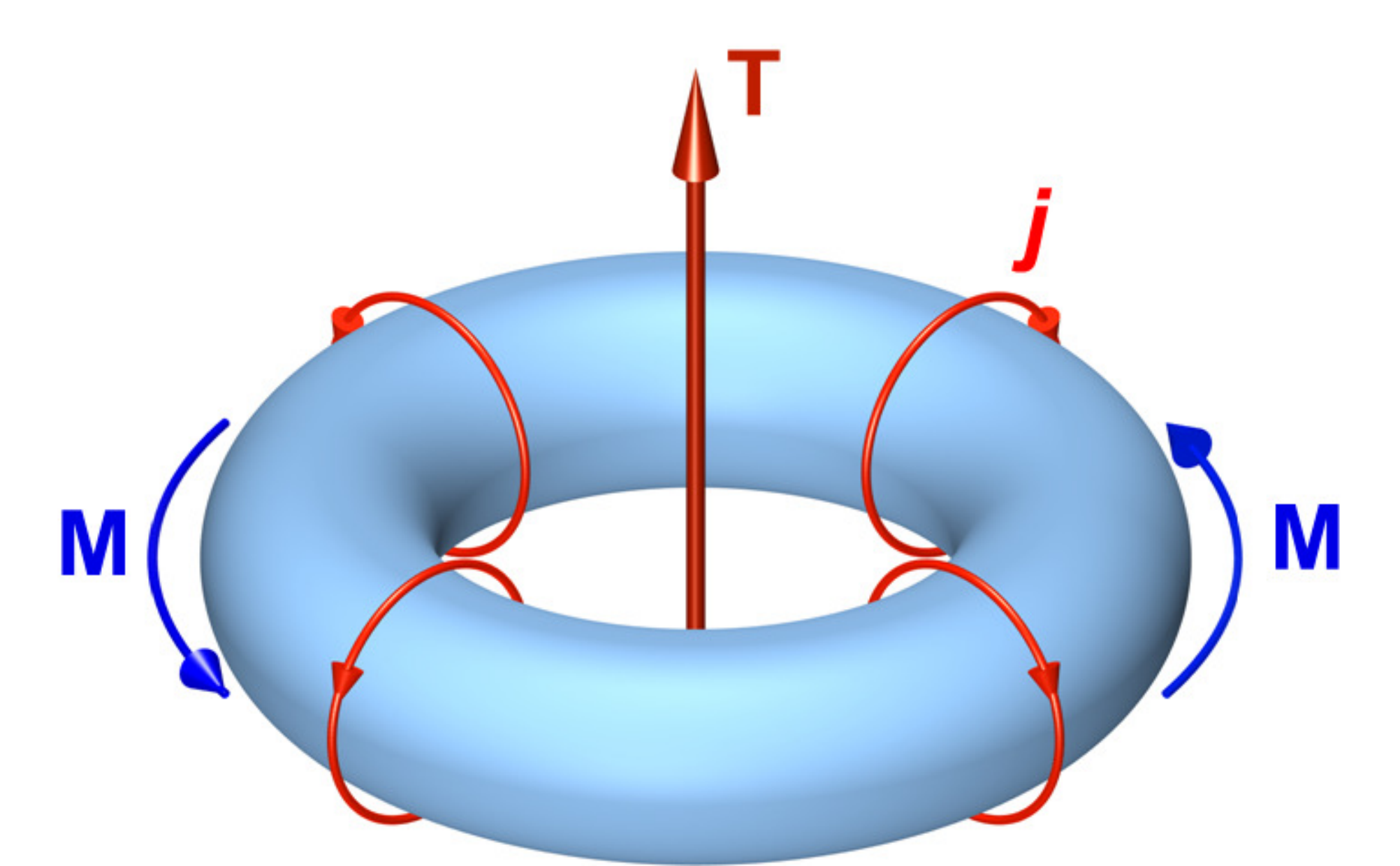}
\captionsetup{justification=justified, format = plain}
\caption{A toroidal dipole moment \textbf{T} generated by the poloidal currents \textbf{\textit{j}} flowing on a torus's surface. It can also be represented as a closed loop of circulating magnetic field \textbf{M}.}
\label{Figure 1}
\end{figure}

Interestingly, in the \emph{dynamic} case, the exploration on oscillating anapole states followed a similar path as its static counterpart. In the late 1960s and early 1970s, Dubovik and Cheshkov first theoretically examined the possibility to connect the quantum description of anapole states with classical electrodynamics \cite{dubovik1967form, dubovik1974multipole}. They found out that, a dynamic nonradiating anapole could not be explicitly described by the conventional multipole theory which only includes electric and magnetic multipoles. A new multipole family -- termed 'toroidal multipoles' -- thus had to be introduced to complete the framework of multipole expansion. The first leading term of this multipole family is the toroidal dipole, as illustrated in Fig. \ref{Figure 1}. By interfering with an electric dipole, an anapole excitation could be achieved, as we will discuss in detail in subsequent sections. The toroidal multipoles, after introduced by Dubovik and Cheshkov, were then investigated in various contexts. For example, in 1985, Ginzburg and Tsytovich \cite{ginzburg1985fields} studied the fields and the Cherenkov radiation of a toroidal dipole moment moving at a constant velocity. Later, a toroidal dipole in arbitrary motion was considered by Heras \cite{heras1998electric}. In 2000, the concept of toroidal polarization or toroidization was introduced into condensed matter physics to describe the density of toroidal dipoles in continuous media, similar to the macroscopic electric polarization and magnetization \cite{dubovik2000material}.  However, despite the increasing theoretical interest and experimental attempts on observing dynamic anapoles and toroidal multipoles \cite{tolstoy1990new,fedotov2007aromagnetism,papasimakis2009gyrotropy}, direct evidence of their existence was still lacking because they are usually very weak and easily masked by the contributions of electric and magnetic multipoles in natural materials. 

The timely emergence and ensuing rapid development of metamaterials offered a natural solution to this conundrum by allowing for an unprecedented control of light-matter interactions. In 2010, Zheludev's group reported the first experimental observation of a dynamic toroidal resonance in a microwave metamaterial \cite{kaelberer2010toroidal}. The constituent split wire loops were intentionally designed to suppress the leading multipoles and highlight their toroidal response. In 2013, the first dynamic anapole excitation was also realized by the same group in the microwave regime \cite{fedotov2013resonant}. Later, they theoretically studied THz toroidal response supported in dielectric metamaterials, which may exhibit significantly lower loss compared to their metallic counterparts \cite{basharin2015dielectric}. In 2015, Miroshnichenko \textit{et al}. pioneered the first study of anapole states in the visible region \cite{miroshnichenko2015nonradiating}. They thoroughly investigated the anapole supported in individual high-index dielectric nanoparticles and successfully observed it by using an extremely simple geometry -- silicon nanodisks. Spurred on by all these great efforts, considerable research interest has been generated in the subject of anapole states and led to many amazing feats in a very short time, particularly in the field of nanophotonics and metamaterials. 

\subsection{Mie scattering of nanoparticles: a seeming paradox}
The study of light scattering has a long history and is still a vibrant research field that continues to surprise with new insights \cite{fan2014light, koenderink2015nanophotonics}. In this regard, it also offers us a transparent window to look at the physical origin of dynamic anapoles. Here we examine the most general case of optical scattering by nanosized particles -- a dielectric nanosphere situated in the vacuum. We can then perform an analytical analysis using classical Mie theory \cite{bohren2008absorption}, which provides an exact solution to the scattering problem. The scattering cross section $Q_{\rm{scat}}$  of the nanosphere can be read as:
\begin{eqnarray}
Q_{\rm{scat}} =  \frac{2\pi}{k^2}\sum_{\ell=1}^{\infty}(2\ell+1)(|a_\ell|^2+|b_\ell|^2), \label{eq1}
\end{eqnarray}
where $k = 2\pi f/c$ is the wave number, $f$ and $c$ are the frequency and the phase velocity of the incident light. The contributed scattering coefficients $a_\ell$ (electric) and $b_\ell$ (magnetic)  can be expressed as:
\begin{eqnarray}
a_\ell = \frac{[D_\ell(nkR)/n + \ell/kR]\psi_\ell(kR)-\psi_{\ell-1}(kR)}{[D_\ell(nkR)/n + \ell/kR]\xi_\ell(kR)-\xi_{\ell-1}(kR)}, \label{eq2}\\
b_\ell = \frac{[nD_\ell(nkR) + \ell/kR]\psi_\ell(kR)-\psi_{\ell-1}(kR)}{[nD_\ell(nkR) + \ell/kR]\xi_\ell(kR)-\xi_{\ell-1}(kR)}, \label{eq3}
\end{eqnarray}
where $n$ and $R$ are the refractive index and the radius of the sphere, respectively. $D_\ell(nkR)$ is defined as $D_\ell(nkR) = \psi_{\ell}'(nkR)/\psi_\ell(nkR)$, with $\psi_\ell(kR)$ and $\xi_\ell(kR)$ the Riccati-Bessel functions of the first and second kinds. 

\begin{figure}[h]
\centering
\includegraphics[width=13cm]{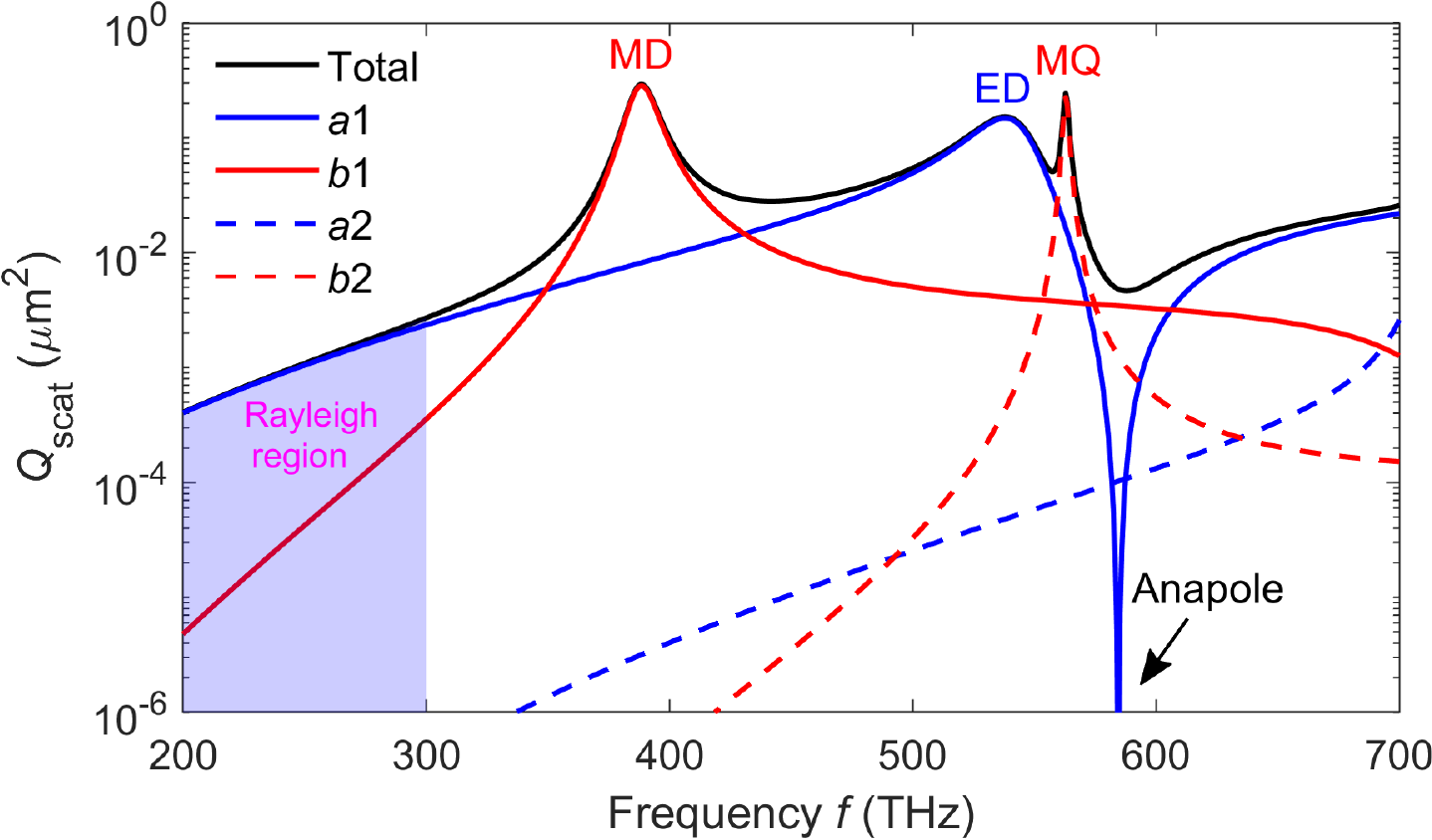}
\captionsetup{justification=justified, format = plain}
\caption{Scattering cross section $Q_{\rm{scat}}$ and its Mie expansion of a dielectric nanosphere with $n = 5$ and $R = 75$ nm. The shaded area represents the Rayleigh region where $R \ll \lambda$. The maxima of the scattering coefficients $a_1$, $b_1$, and $b_2$ correspond to the electric dipole (ED), magnetic dipole (MD), and magnetic quadrupole (MQ) resonances of the nanosphere, respecivetly, whereas the vanishing  state of $a_1$ at $f = 584$ THz ($\lambda = 513$ nm) is associated with the anapole.}
\label{Figure 2}
\end{figure}

In Fig. \ref{Figure 2} we plot the scattering response of a dielectric nanosphere with $n = 5$ and $R = 75$ nm. As we can see, when the particle size is much smaller than the incident wavelength, its scattering response is dominated by the electric dipole term $a_1$ and satisfies the quasistatic (Rayleigh) approximation, following the well-known fourth-power dependence on the frequency of light.  It implies that the electric dipolar polarization should always exist within small particles and its strength increases with the frequency. This effect can be used to explain a variety of interesting physical phenomena, such as the blue color of the sky and the reddish afterglow. 

For higher frequencies, the geometric size and retardation effects result in the breakdown of the above approximation. The incident light would polarize the \emph{bounded} electrons of the dielectric nanoparticle and generate displacement currents, which further lead to a series of optical resonances due to their various distributions. This process is somehow an analog to the localized surface plasmon resonances (LSPRs) of metallic nanospheres, where \emph{free} electrons are oscillating at the frequency of the incident light \cite{baranov2017all}. In our case, the first order optical resonance of the dielectric nanosphere corresponds to the maximum of the scattering coefficient $b_1$, indicating a magnetic dipole (MD) resonance. This is due to the high refractive index of the particle, which produces a circulating displacement current when $f_{\rm{MD}} \approx c/2nR$. Similarly, the second order resonance is associated with the maximum of the scattering coefficient $a_1$, implying an electric dipole (ED) resonance due to the strongest electric dipolar polarization. When the frequency continues increasing, a series of higher-order resonances such as the magnetic quadrupole (MQ) resonance would appear subsequently. Such optically-induced Mie resonances of high-index nanoparticles and associated dynamic magnetism are now on the spotlight of research in nanophotonics. We refer the readers to several recent reviews covering many different aspects of this subject \cite{jahani2016all, smirnova2016multipolar, kuznetsov2016optically, liu2017multipolar, staude2017metamaterial, kruk2017functional, baranov2017all, kivshar2018all, liu2018generalized}

Besides the above set of optical resonances which dramatically enhance the scattering response of the nanosphere, there is also a noteworthy spectral position at $f = 584$ THz ($\lambda = 513$ nm), where the scattering coefficient $a_1$ is completely vanishing with an accompanying dip in the total scattering response. Since the scattering coefficient $a_1$ usually represents the electric dipolar polarization within the particle, the absence of $a_1$ seemingly suggests that the particle is not even polarized. This conclusion apparently contradicts with our \textit{a priori} knowledge. On the other hand, Mie theory also allows us to calculate the electric fields inside the nanosphere with the internal coefficient $d_\ell$ \cite{bohren2008absorption}:
\begin{eqnarray}
d_\ell = \frac{n\psi_\ell(kR)\xi_{\ell}'(kR)-n\psi_{\ell}'(kR)\xi_{\ell}(kR)} {n\psi_\ell(nkR)\xi_{\ell}'(kR)-\psi_{\ell}'(nkR)\xi_{\ell}(kR)}, \label{eq4}
\end{eqnarray}
where $d_\ell$ correlates with the scattering coefficient $a_\ell$ via the transverse boundary conditions which can be written in the following form:
\begin{eqnarray}
\psi_\ell(nkR)d_\ell + \xi_\ell(kR)a_\ell = \psi_\ell(kR), \label{eq5} \\
\psi_{\ell}'(nkR)d_\ell + n\xi_{ell}'(kR)a_\ell = n\psi_{\ell}'(kR). \label{eq6}
\end{eqnarray}

\begin{figure}[h]
\centering
\includegraphics[width=13cm]{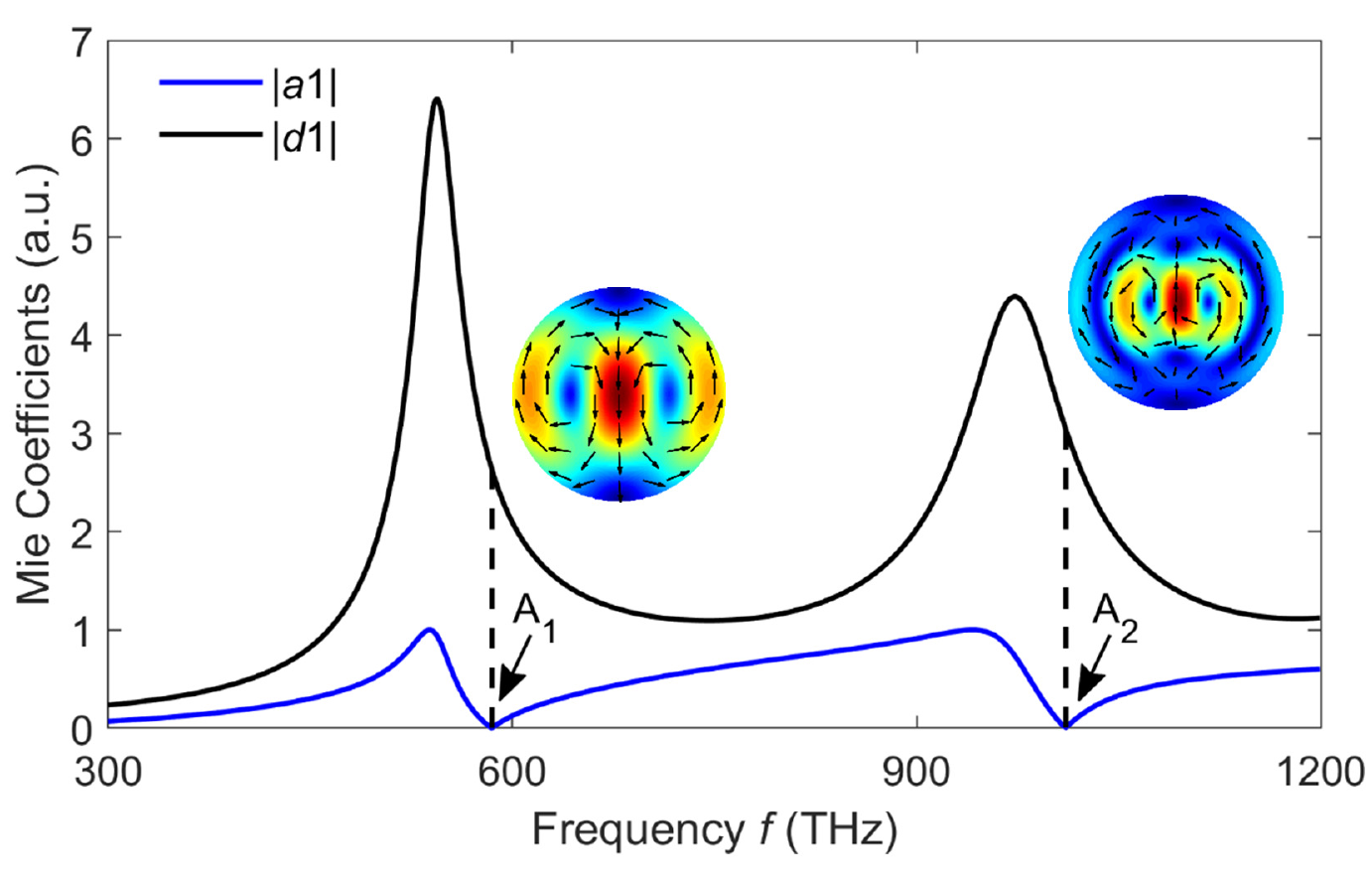}
\captionsetup{justification=justified, format = plain}
\caption{Spectral response of the Mie scattering and internal coefficients $a_1$ and $d_1$. A$_1$ and A$_2$ denote the first-order and second-order anapole states when $a_1 \approx 0$ but $d_1 > 1$. The insets show the electric field distributions of the two anapole states A$_1$ and A$_2$. The black arrows represent electric field vectors.}
\label{Figure 3}
\end{figure}

The amplitudes of the scattering and internal Mie coefficients $a_1$ and $d_1$ are depicted in Fig. \ref{Figure 3}. It can be easily observed that, when $a_1$ vanishes at $f = 584$ THz, the amplitude of $d_1$ is still larger than 1, which indicates a non-zero electric dipolar excitation inside the nanosphere. However, this time-varying distribution of electric dipoles does not radiate into the far field as its corresponding scattering coefficient $a_1$ just vanishes! The insets further give the exact electric-field profiles at the A$_1$ state, where we can clearly see how the nontrivial, yet nonradiating fields distribute inside the sphere. The prominent poloidal currents circulating around the sphere give rise to a dynamic field configuration similar to that of the static anapole proposed by Zeldovich \cite{zel1958electromagnetic}. Therefore, we call this peculiar spectral position as an anapole state. For lossless dielectric nanoparticles, the suppression of far-field scattering at anapole states also accompanies with enhanced near fields, which can facilitate many near-field effects. In addition, the presence of higher-order anapole states can be also seen in a broader frequency range (or equivalently with larger particle sizes). Such a higher-order anapole state A$_2$ can be treated as a hybridization of the first-order anapole state A$_1$ and a Fabry-Perot resonance, which enables a stronger localization of near-field energy \cite{yang2017multimode}. Recently, the exotic feature of higher-order anapole states was thoroughly investigated by Zenin \textit{et al.} \cite{zenin2017direct} using near-field optical microscopy (SNOM) and has been utilized in multispectral field enhancements \cite{yang2018anapole} and nonlinear optics \cite{grinblat2016efficient}. 

\subsection{Toroidal multipoles and nonradiating anapole states}
To resolve the seeming paradox concealed in Mie theory, we then adopt a more general multipole expansion approach to study the radiation fields created by arbitrary charge-current distributions. We assume that there exist localized distributions of prescribed charge $\rho(\mathbf{r})$ and current $\mathbf{J}(\mathbf{r})$ with harmonic time dependence $\rme^{-\rmi\omega t}$. Given the continuity equation $\rmi \omega \rho(\mathbf{r}) = \nabla\cdot\mathbf{J(r)}$. The Maxwell equations can be written as:
\begin{eqnarray}
\boldsymbol{\nabla}\cdot\mathbf{E'(r)} = 0, \label{eq7}\\
\boldsymbol{\nabla}\cdot\mathbf{H'(r)} = 0,  \label{eq8}\\
\boldsymbol{\nabla}\times\mathbf{E'(r)} = \rmi kZ_0\mathbf{H'(r)} +  \frac{\rmi}{\omega\epsilon_0}\nabla\times\mathbf{J(r)}, \label{eq9}\\
\boldsymbol{\nabla}\times\mathbf{H'(r)} = -\rmi k\mathbf{E'(r)}/Z_0, \label{eq10}
\end{eqnarray}
where $k$ is the wave number; $\epsilon_0$ and $Z_0$ are the permittivity and the impedance of free space. Here we define the divergenceless electric field $\mathbf{E'(r)}$ as $\mathbf{E'(r)} = \mathbf{E} + \frac{\rmi}{\omega\epsilon_0} \mathbf{J(r)}$ and magnetic field $\mathbf{H'(r)} = \mathbf{B(r)}/\mu_0$, following the treatment in Jackson's book \cite{jackson_classical_1999}. We can then obtain the inhomogeneous wave equations for the scalars $\mathbf{r} \cdot \mathbf{E'(r)}$ and $\mathbf{r} \cdot \mathbf{H'(r)}$ :
\begin{eqnarray}
(\nabla^2 + k^2)\mathbf{r}\cdot\mathbf{E'(r)} = \frac{Z_0}{k}\mathcal{L}\cdot \big[\boldsymbol{\nabla}\times\mathbf{J(r)} \big], \label{eq11}\\
(\nabla^2 + k^2)\mathbf{r}\cdot\mathbf{H'(r)} = -\rmi \mathcal{L} \cdot \mathbf{J(r)}, \label{eq12}
\end{eqnarray}
with the angular momentum operator $\mathcal{L} = -\rmi\mathbf{r}\times \boldsymbol{\nabla}.$ These could be solved by using the scalar Green's function of the Helmholtz equation:
\begin{eqnarray}
(\nabla^2 + k^2)G(\mathbf{r},\mathbf{r'}) = -\delta(\mathbf{r} - \mathbf{r'}), \label{eq13}\\
G(\mathbf{r},\mathbf{r'}) = \frac{\rme^{\rmi k |\mathbf{r'}-\mathbf{r}}|}{4\pi|\mathbf{r}-\mathbf{r'}|}. \label{eq14}
\end{eqnarray}
The exact expansion of the Green's function into spherical harmonics is given by:
\begin{eqnarray}
G(\mathbf{r},\mathbf{r'})  = \rmi k \sum_{\ell=0}^{\infty}\sum_{m=-\ell}^{\ell}j_\ell(kr')h_\ell^{(1)}(kr)Y_{\ell,m}(\theta,\phi)Y_{\ell,m}^*(\theta',\phi'). \label{eq15}
\end{eqnarray}
where $j_\ell$ and $h_\ell^{(1)}$ are the spherical Bessel and Hankel functions of the first kind, respectively. $Y_{\ell,m}$ is the scalar spherical harmonics. By substituting the expanded Green's function into Eqs. (\ref{eq11}) and (\ref{eq12}), then we can fully decompose the radiation fields in terms of the so-called Hansen multipoles \cite{jackson_classical_1999, hansen1935new}:
\begin{eqnarray}
\mathbf{E'(r)} = \sum_{\ell,m}\Big[ E_{\ell,m}\boldsymbol{\Psi}_{\ell,m} + M_{\ell,m}\boldsymbol{\Phi}_{\ell,m} \Big], \label{eq16}\\
\mathbf{H'(r)} = \sum_{\ell,m}\Big[ E_{\ell,m}\boldsymbol{\Phi}_{\ell,m} - M_{\ell,m}\boldsymbol{\Psi}_{\ell,m}  \Big]/Z_0, \label{eq17}
\end{eqnarray}
where $\boldsymbol{\Phi}_{\ell,m}$ and $\boldsymbol{\Psi}_{\ell,m}$ represent the vector spherical harmonics with the following definitions:
\begin{eqnarray}
\boldsymbol{\Phi}_{\ell,m} = \frac{h_\ell^{(1)}(kr)}{\sqrt{\ell(\ell+1)}}\mathcal{L}Y_{\ell,m}(\theta,\phi), \label{eq18} \\
\boldsymbol{\Psi}_{\ell,m} = \frac{\rmi}{k}\boldsymbol{\nabla}\times\boldsymbol{\Phi}_{\ell,m}, \label{eq19}
\end{eqnarray}
and determine the expansion coefficients $E_{\ell,m}$ and $H_{\ell,m}$ by:
\begin{eqnarray}
E_{\ell, m} = \frac{Z_0k^2}{\rmi \sqrt{\ell(\ell+1)}}\int Y_{\ell, m}^{*} \Big[c\rho(\mathbf{r}) \frac{\partial}{\partial r}[r j_\ell(kr)] + \rmi k [\mathbf{r} \cdot \mathbf{J(r)}]  j_\ell(kr) \Big]\rmd^3r, \label{eq20} \\
M_{\ell, m} =  \frac{Z_0k^2}{\rmi \sqrt{\ell(\ell+1)}}\int Y_{\ell, m}^{*} \nabla \cdot [\mathbf{r} \times \mathbf{J(r)}] j_{\ell}(kr) \rmd^3 r. \label{21}
\end{eqnarray}
Note that the coefficient $E_{\ell,m}$ can be further split into two terms $Q_{\ell,m}$ and $T_{\ell,m}$:
\begin{eqnarray}
E_{\ell,m} = Q_{\ell,m} + T_{\ell,m}, \label{eq22} \\
Q_{\ell,m} = \frac{Z_0k^2}{\rmi \sqrt{\ell(\ell+1)}}\int Y_{\ell, m}^{*}c \frac{\partial}{\partial r}[r j_\ell(kr)] \rho(\mathbf{r}) \rmd^3r, \label{eq23} \\
T_{\ell,m} = \frac{Z_0k^2}{\sqrt{\ell(\ell+1)}}\int k Y_{\ell, m}^{*}  j_\ell(kr) [\mathbf{r} \cdot \mathbf{J(r)}] \rmd^3r, \label{eq24}
\end{eqnarray}
which results in three types of expansion coefficients $Q_{\ell,m}$, $M_{\ell, m}$, and $T_{\ell,m}$. It is worth noting that these multipole coefficients are directed linked to three types of distinct charge-current distributions: $Q_{\ell,m}$ is dependent on the distribution of charge density $\mathbf{\rho(\mathbf{r})}$; $M_{\ell,m}$ is generated by the transverse currents ($\mathbf{r}\times \mathbf{J} \neq 0$) while $T_{\ell,m}$ is dependent on the radial currents ($\mathbf{r}\cdot\mathbf{J} \neq 0$). Therefore, they are often referred to as electric ($Q_{\ell,m}$), magnetic ($M_{\ell,m}$), and toroidal ($T_{\ell,m}$) multipole moments, characterizing the strength of each dynamic multipole, respectively \cite{dubovik1990toroid, radescu2002exact, radescu2002toroid}.
All together, we obtain the radiated electric and magnetic fields of an arbitrary charge-current distribution as \cite{papasimakis2016electromagnetic}:
\begin{eqnarray}
\mathbf{E'(r)} = \sum_{\ell,m}\Big[ Q_{\ell,m}\boldsymbol{\Psi}_{\ell,m} + M_{\ell,m}\boldsymbol{\Phi}_{\ell,m} +  T_{\ell,m}\boldsymbol{\Psi}_{\ell,m} \Big], \label{eq25}\\
\mathbf{H'(r)} = \sum_{\ell,m}\Big[ Q_{\ell,m}\boldsymbol{\Phi}_{\ell,m} - M_{\ell,m}\boldsymbol{\Psi}_{\ell,m} +  T_{\ell,m}\boldsymbol{\Phi}_{\ell,m}  \Big]/Z_0. \label{eq26}
\end{eqnarray}
Now we can directly compare the charge-current expansion with Mie theory. First of all, both methods provide a complete set of solutions to scattering problems. However, while there are three types of multipole families in the above expansion Eqs.(\ref{eq25}, \ref{eq26}), only two scattering coefficients $a_\ell$ and $b_\ell$ are present in Mie theory. This significant difference lies in the fact that the coefficients in Mie theory are determined by the boundary conditions only respected for the transverse field components \cite{bohren2008absorption}. Thus, the two kinds of Mie scattering coefficients $a_\ell$ and $b_\ell$ are not directly related to any specific charge-current distributions but only dependent on the spatial distributions (TE or TM) of the scattered fields. Since the electric and toroidal multipole moments both exhibit TM-type radiation ($Q_{\ell,m}$ and $T_{\ell,m}$ share the same spherical harmonics $\boldsymbol{\Psi}_{\ell,m}$), their individual contributions to the total scattering are emerged into one single term $a_\ell$ in Mie theory and thereby cannot be separately identified. By contrast, electric and toroidal multipoles are well defined in the charge-current expansion in terms of their different physical origins. Thus, we can write down the \emph{generalized conditions} for anapole states as:
\begin{eqnarray} 
\it{Q}_{\ell,m} = - \it{T}_{\ell, m}. \label{eq27}
\end{eqnarray}

\begin{figure}[b]
\centering
\includegraphics[width=13cm]{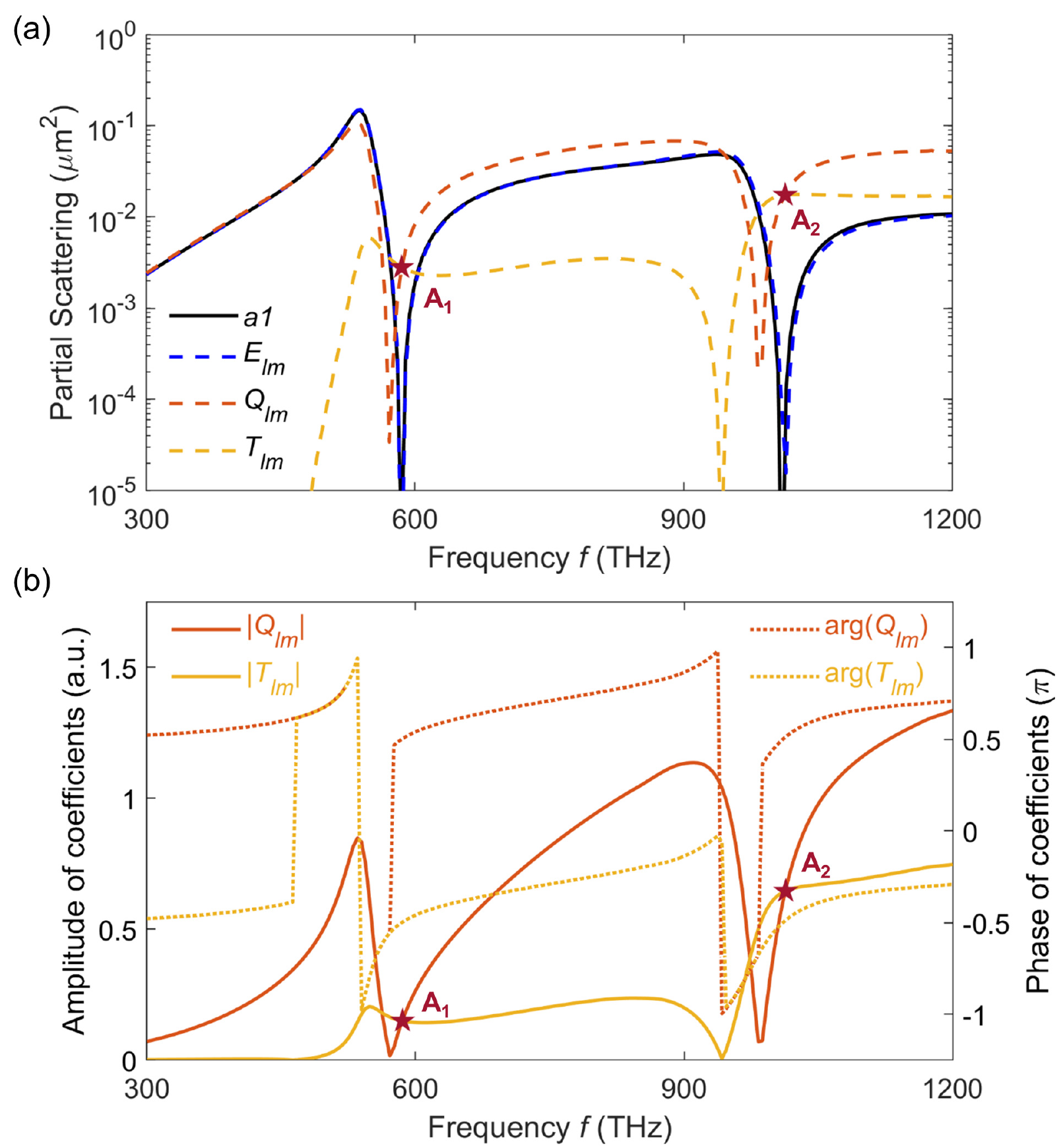}
\captionsetup{justification=justified, format = plain}
\caption{Comparison between Mie theory and charge-current expansion. (a) Partial scattering from Mie coefficients $\it{a}_1$ and charge-current expansion coeffcients $\it{E}_{11}$, $\it{Q}_{11}$, and $\it{T}_{11}$. The star symbols represent the spectral positions when $|\it{Q}_{11}| =|\it{T}_{11}|$, which are consistent with $\it{a}_1 \to 0$ in Mie theory. (b) Spectral amplitude and phase plots of the coefficients $\it{Q}_{11}$ and $\it{T}_{11}$, verifying the validity of the derived condition $\it{Q}_{lm} =- \it{T}_{lm}.$ }
\label{Figure 4}
\end{figure}

To verify the above condition, let's move back to our previous example of the dielectric nanosphere. In Fig. \ref{Figure 4}a we plot the partial scattering contributed by the first leading coefficients ($\ell = 1$) in Mie theory and the charge-current approach. Given the spherical symmetry, here we only consider the case of $m = 1$. It is clearly seen that the charge-current coefficient $\it{E}_{11}$ is equivalent to the Mie coefficient $\it{a}_1$, both of which fully determine the TM$_{11}$-type (electric-dipole-like) far-field scattering of the nanosphere. The vanishing of these two coefficients defines the spectral positions of the anapole states, of which the first two orders are A$_1$ ($f$ = 584 THz) and A$_2$ ($f$ = 1010 THz), respectively. The spectral positions of A$_1$ and A$_2$ also coincide with the anti-phase intersections of $\it{Q}_{\ell m}$ and $\it{T}_{\ell m}$, as marked by the star symbols in Fig. \ref{Figure 4}b.  

So far we have now fully examined and linked the appearance of anapole states in two representative expansion frameworks which are often used in nanophotonics research. We set up a solid foundation for understanding the physical origins and generalized excitation conditions of anapole states by virtue of charge-current expansion. It is worth noting that, the existence of ideally isolated toroidal multipoles and their independence from electric multipoles are still interesting topics on debate \cite{papasimakis2016electromagnetic, fernandez2017dynamic, alaee2018electromagnetic, talebi2018theory}. However, since anapole states could be identified unambiguously in both expansion approaches, the nature of toroidal multipoles, i.e. whether they represent an independent multipole family, or they are essentially higher-order electric multipoles beyond long-wavelength approximation, would not influence the generation and intrinsic properties of anapole states. As a final remark in this section, we mention that anapole states could also be characterized via several other theoretical pathways, such as quasinormal modes \cite{powell2017interference} or Fano-Feshbach projection schemes \cite{gongora2017fundamental}. In addition, the concept of anapole or nonradiating states could also be extended to higher-order electric multipole \cite{li2018origin} (e.g. $l = 2$ in Eq. \ref{eq27}) or magnetic multipole families \cite{luk2017hybrid, yang2017multimode}.

\section{Applications}

In the previous section, we have explicitly showed the general existence and formation mechanisms of anapole states. Their experimental realization, however, is elusive as they are usually spectrally overlapped with other radiating multipoles. Note that the toroidal term $\it{T}_{\ell m}$ in Eq. (\ref{eq24}) scales with an extra $k$-dependence compared to its electric and magnetic counterparts, indicating that the partial scattering by toroidal family is usually much weaker. This is also partly the reason why anapole states are often neglected in conventional descriptions of multipole expansion. 

In this regard, nanophotonics and metamaterials offer us unprecedented degrees of freedom to manipulate the interaction between electromagnetic waves and structured matters. This has enabled us to explore many new ways to achieve and exploit dynamic anapoles. Here, we will discuss recent advances in the understanding and applications of anapole states and highlight some of the exciting directions for this field.

\subsection{Nonradiating sources and tailored light scattering}

\begin{figure}[t]
\centering
\includegraphics[width=15.5cm]{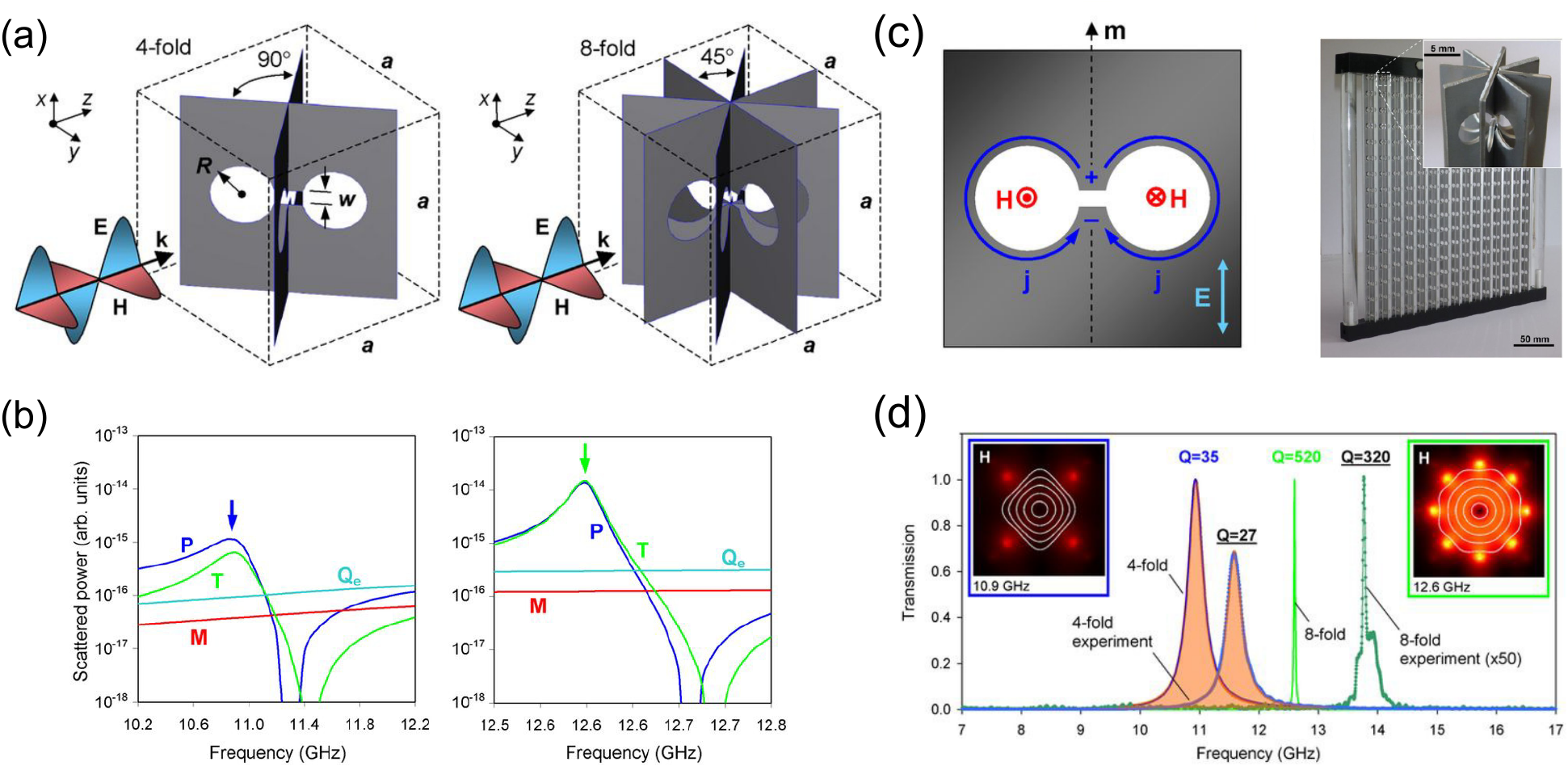}
\captionsetup{justification=justified, format = plain}
\caption{First experimental observation of dynamic anapole states in microwave metamaterials. 
(a) Schematics of constituent meta-atoms with 4-fold (left) and 8-fold (right) symmetry. 
(b) Corresponding multipolar scattering of the two kinds of meta-atoms. (c) Induced charge-current distributions around a dumbbell-shaped aperture (left) and a photograph of the constructed stainless steel slab (right). $\mathbf{m}$ represents the axis of the mirror symmetry. (d) Simulated (solid lines) and experimental (dotted lines) transmission spectra of the anapole metamaterials. Green and blue curves represent the response of 4-fold and 8-fold symmetric metamaterials, respectively. Insets show simulated magnetic field lines and intensity distributions of 4-fold (left) and 8-fold (right) symmetric meta-atoms. Reprinted from \cite{fedotov2013resonant}. CC BY 3.0.}
\label{Figure 5}
\end{figure}

To realize a strong anapole excitation, the most important question is how to increase the toroidal response of a material while suppressing its conventionally dominant electric and magnetic responses. In 2013, Fedotov \emph{et al}. \cite{fedotov2013resonant} designed an elegant microwave metamaterial with enhanced toroidal dipole resonance and successfully observed its destructive interference with a collocated electric dipole, hence for the first time experimentally realizing a dynamic anapole state in the electromagnetic spectrum. They etched dumbbell-shaped apertures in a metal screen made of stainless steel, as shown in Fig. \ref{Figure 5}(a). Two different metamaterial configurations consisting of 4-fold and 8-fold elements were proposed to mimic a rotational symmetry of a torus. When it interacts with an impinging wave, the induced poloidal currents along the edges of the circular cuts would circulate along the meridians of the metamaterial, thus giving rise to a strong toroidal dipole moment $\bf{T}$. Meanwhile, the shared waist of all the apertures in the center of the meta-atoms contributes to an electric dipole response $\bf{P}$, as shown in Fig. \ref{Figure 5}(c). By adjusting related geometric parameters of the apertures, a balance between $\bf{P}$ and $\bf{T}$ could be optimized at desired frequencies. As in Fig. \ref{Figure 5}(b), by virtue of Cartesian multipole expansion, it can be clearly seen that the strength of the toroidal response $\bf{T}$ is at the same level as that of $\bf{P}$ for both kinds of meta-atoms. In Fig. \ref{Figure 5}(d), two prominent peaks in the transmission spectra of the metamaterials indicate the appearance of an interesting phenomenon: anapole-induced transparency. For designs with 4-fold and 8-fold symmetry, significantly high Q-factors of 27 and 320 were experimentally achieved, respectively. The simulated magnetic field intensity distributions with closed loops of magnetic field lines further confirm the presence of anapole excitation inside the mata-atoms.

The first experimental success in the microwave regime undoubtedly revealed the attractive prospect of dynamic anapoles and stimulated a number of relevant studies \cite{guo2014electric, li2014resonant, li2015low, zhang2015optical}. However, such design is not easy to be directly shrunk down to the nanometer scale and efficiently work in higher frequencies as metallic structures would possess strong plasmonic resonances with high Ohmic losses, which largely influence anapole generation and its nonradiating characteristics.  Moreover, in order to enhance toroidal response, complex structures such as vertical dumbbell apertures are also very challenging to realize with high precision using standard nanofabrication techniques. 

\begin{figure}[t]
\centering
\includegraphics[width=15.5cm]{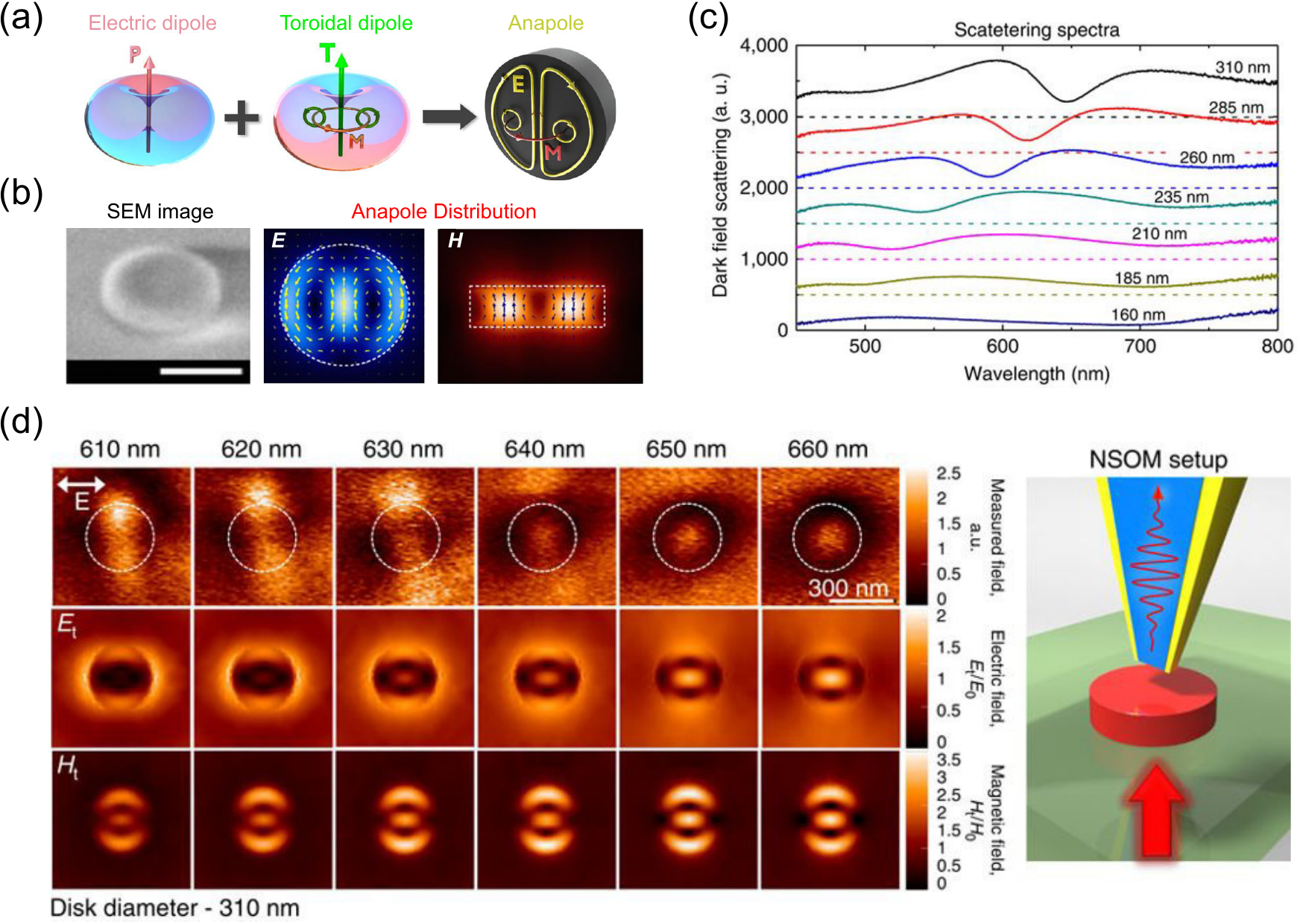}
\captionsetup{justification=justified, format = plain}
\caption{First experimental observation of dynamic anapole states in individual nanoparticles. (a) Illustrative mechanism of anapole excitation in nanodisks. (b) SEM image of a fabricated silicon nanodisk and its simulated electric and magnetic field distributions at anapole states. The scale bar in the SEM image represents 200 nm. (c) Experimental dark-field scattering spectra of silicon disks with a fixed height of 50 nm and a varied diameter ranging from 160 nm to 310 nm. (d) Experimental (top) and simulated (middle: transverse electric; bottom: transverse magnetic) near-field maps of the silicon nanodisk with a diameter of 310 nm. The right panel shows the schematic of SNOM measurement with substrate illumination and a metal-coated tip. Reprinted from \cite{miroshnichenko2015nonradiating}. CC BY 4.0.}
\label{Figure 6}
\end{figure}

To tackle this conundrum, in 2015, Miroshnichenko \emph{et al.} \cite{miroshnichenko2015nonradiating} utilized a new physical platform, i.e. high-index dielectrics, to fundamentally examine anapole states in \emph{individual} nanostructures.  In contrast to their metallic counterparts where the electric-type response is usually dominant in simple geometries, high-index dielectric nanoparticles can support a series of multipole Mie resonances with distinct physical features. By shaping a dielectric nanoparticle from a sphere to a disk or a rectangular patch, one can easily control the spectral positions and relative strengths of all these multipole resonances and its interferences, therefore offering more degrees of freedom to tailor the optical resonances \cite{staude2013tailoring, yang2015controlling, ee2015shape}. On the basis of this notion, subwavelength-thin silicon nanodisks with a large diameter were dedicatedly designed, as shown in Fig. \ref{Figure 6}. In this way, the small heights of the disks would cause dramatic blue shifts of the MD resonance and move it out of the spectral region of interest, while a large diameter-to-thickness aspect ratio facilitates the generation of a toroidal current along the meridian of the disk. Thus, the signature of anapole states in the silicon nanodisks could be easily identified in far-field measurement by pronounced dips in the scattering spectra (Fig. \ref{Figure 6}c). The spectral positions of the anapoles also redshift with an increasing diameter. SNOM was further used to observe the near-field features of the anapole states, as depicted Fig. 6(d). For the disk with a diameter of 310 nm, a hotspot with maximum near-fields appear in the middle of the disk at the anapole wavelength $\sim$ 650 nm, which correlates well with the simulation results. 

\begin{figure}[h!]
\centering
\includegraphics[width=15cm]{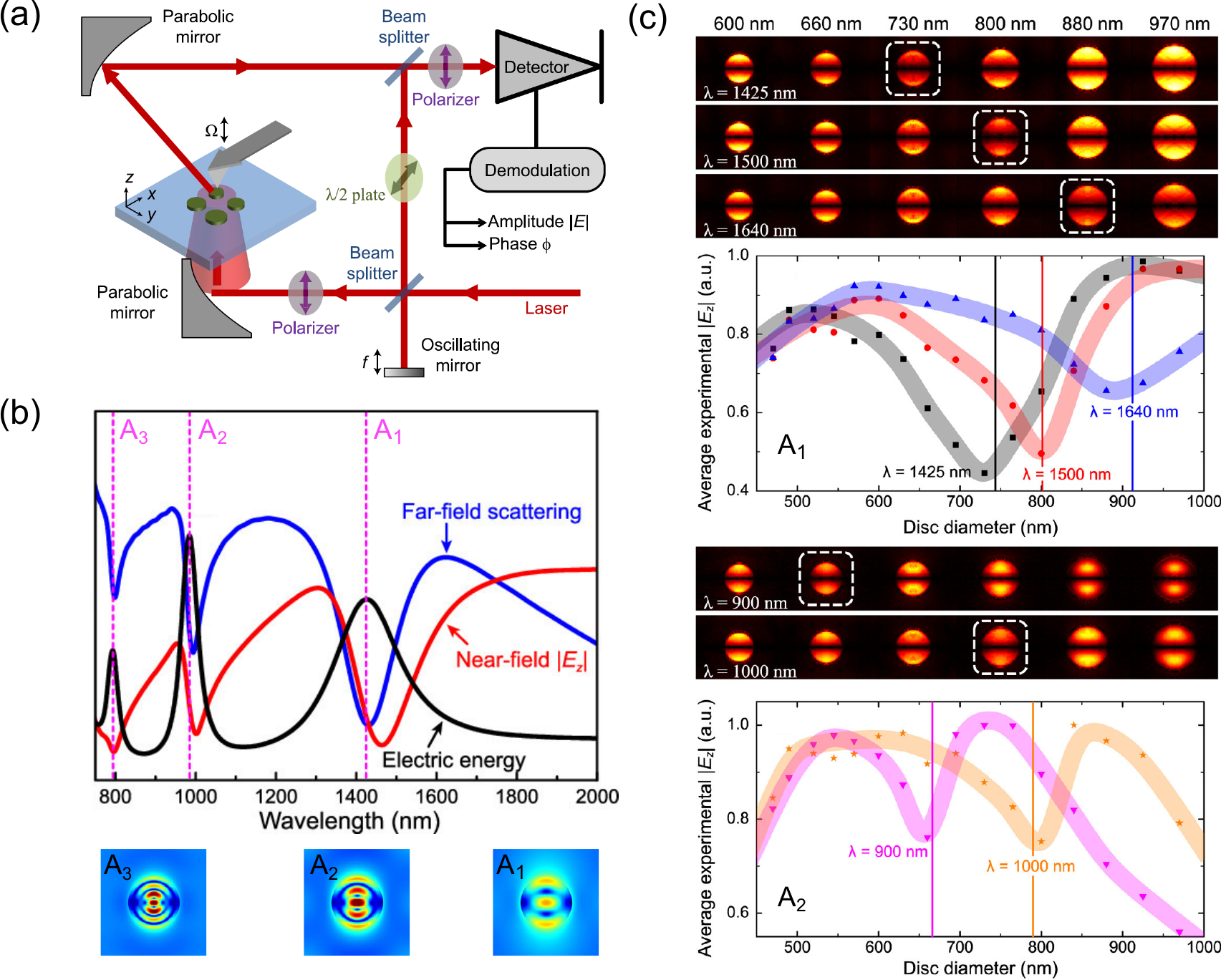}
\captionsetup{justification=justified, format = plain}
\caption{Direct near-field detection of higher-order anapole states. (a)  SNOM setup with interferometric pseudoheterodyne detection. (b) Evidence of anapole states in three different identification approaches, i.e., suppression in far-field scattering and near-field normal component $|E_z|$ as well as enhancement in near-field electric energy. Here a silicon nanodisk in the vacuum with an 80-nm height and an 800-nm diameter is used as an illustrative example. The bottom figures show the electric-field intensity distribution of corresponding anapole states $A_1$, $A_2$, and $A_3$. (c) Experimental near-field $|E_z|$ distribution and its average value for silicon disks with varying diameters, measured at different illumination wavelengths. The disks with the lowest $|E_z|$ are encircled to highlight the corresponding anapole states. The up and bottom plots are related to $A_1$ and $A_2$ states, respectively. Reprinted with permission from \cite{zenin2017direct}. Copyright (2017) American Chemical Society.}
\label{Figure 7}
\end{figure}

The pioneering work in silicon nanodisks reported the first experimental observation of anapole states in the visible region. More importantly, it uncovered a convenient new way to obtain and exploit anapole excitation in the nanophotonics regime. This encouraging finding immediately spurred a variety of investigations on anapole states in different nanophotonic systems. Among all these developments, one of the most straightforward applications of anapole states is to controllably suppress far-field scattering and render transparency to nanoparticles. In this context, Liu, Luk'yanchuk, and Basharin \emph{et al.} demonstrated that the nonradiating feature of anapole states can naturally provide invisibility to nanospheres with radial anisotropy, nanowires, core-shell nanoparticles, and cylinder matrices \cite{liu2015elusive, liu2015toroidal, liu2015invisible, luk2017suppression, nemkov2017nontrivial, ospanova2018multipolar}. Besides the suppression in total scattering, the presence of anapole states can also serve as an effective 'eraser' to nullify scattering contribution from specific multipolar terms. For instance, by utilizing anapole states to cancel electric dipole scattering, Feng and Xu \emph{et al.} showed that a nonmagnetic coreshell nanoparticle can scatter light as a pure magnetic dipole \cite{feng2017ideal}. Meanwhile, the anapole states have also been used to enhance absorption in dielectric materials with moderate losses \cite{wang2016engineering} and to provide viable platforms for observing new physical phenomena such as the time-dependent Aharonov-Bohm effect \cite{nemkov2017nonradiating, savinov2018light}.

Besides the fundamental anapole, it is also of high importance to explore higher-order anapole states as they possess a stronger concentration of near-field energy and narrower spectral responses. In 2017, Zenin \emph{et al.} thoroughly investigated the signatures of first- and higher-order anapole states in different identification approaches \cite{zenin2017direct}, as shown in Fig. \ref{Figure 7}. In particular, they proposed a novel mechanism to ambiguously identify anapole states of different orders by a near-field detection method with resolved amplitude and phase information. They found that the anapole states feature noticeable local minima of the normal component $|E_z|$ in the near field. As shown in Fig. \ref{Figure 7}(c), silicon nanodisks with a varied diameter from 600 to 970 nm were examined by SNOM at five different wavelengths. For each wavelength, a clear drop in $|E_z|$ could be observed at a certain disk size, which in turn depends on the wavelength. Further simulations confirm that these drops exactly correlate with the first- and second-order anapole generation and are robust to different altitudes on top of the disks. The presented near-field identification technique could be a particularly useful tool for nanoparticle arrays with subwavelength variations, where far-field techniques might fail due to a resolution limit.

\begin{figure}[t]
\centering
\includegraphics[width=15.5cm]{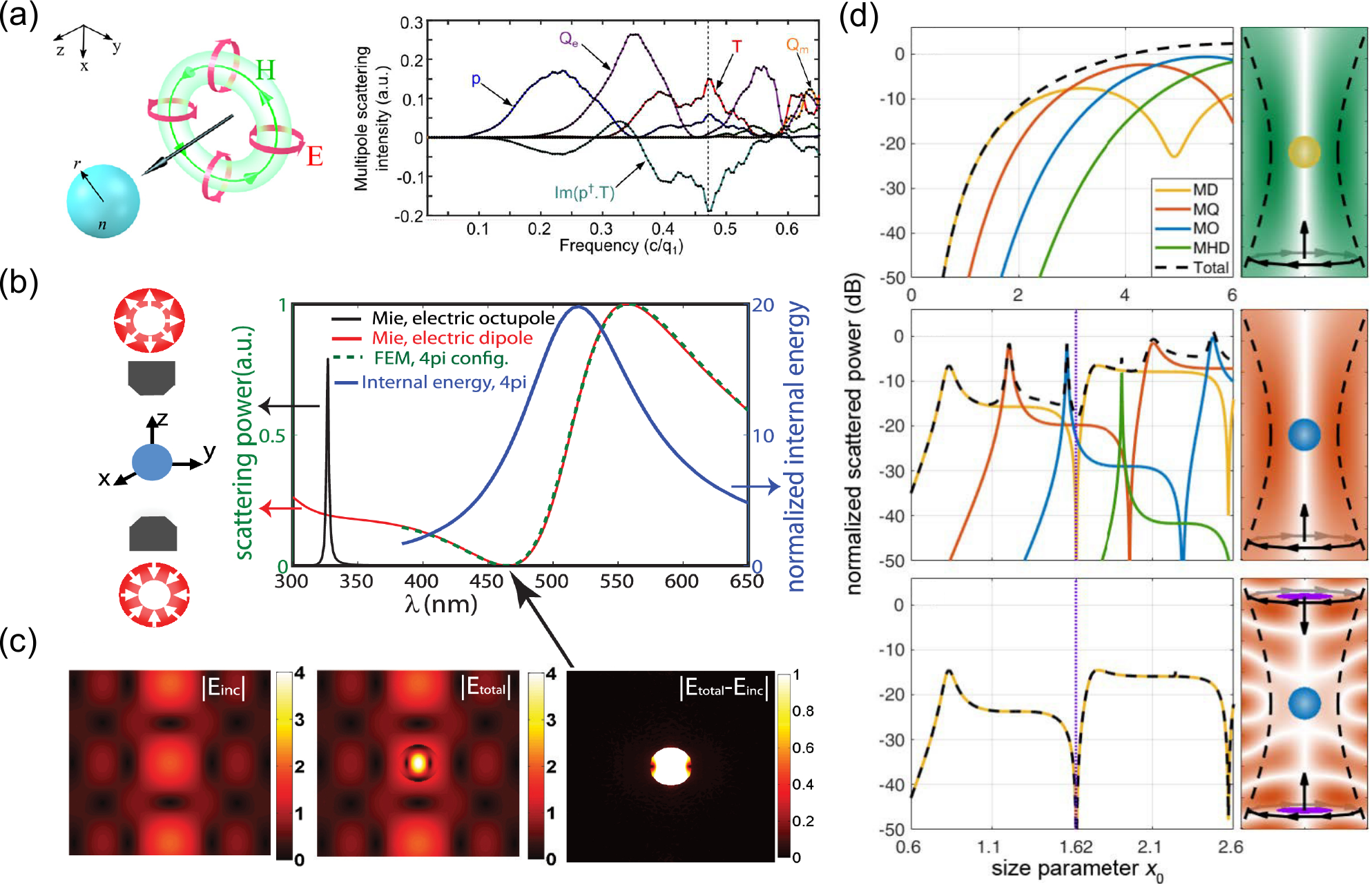}
\captionsetup{justification=justified, format = plain}
\caption{Radiationless anapole states excited by structured light. (a) A dielectric nanosphere ($n = 2$) illuminated by a flying doughnut pulses (left) and its multipolar scattering intensity (right), where $\rm{p,}\ \rm{Q_m,}\ \rm{T,}\ \rm{Q_e}$ represent its electric dipole, magnetic dipole, toroidal dipole, and electric quadrupole moments, respectively. (b) High-index dielectric nanosphere ($n = 3.5$) under 4$\pi$ illumination with radially polarized light (left) and its spectral response of internal energy and scattering power (right). (c) Electric fields of the focal 4$\pi$ illumination (left, background), the particle under the excitation (middle, total field), and the subtracted field (right, induced field). (d) Gold (top) and silicon (middle) nanospheres excited by azimuthally polarized light. The bottom panel shows a silicon nanosphere illuminated by two counter-propagating azimuthally polarized light. The purple dotted line indicate the position of the first-order anapole. (a) Reprinted with permission from \cite{raybould2017exciting}. Copyright (2017), AIP Publishing LLC. (b,c) Reprinted with permission from \cite{wei2016excitation}. Copyright (2016) Optical Society of America. (d) Reprinted with permission from  \cite{lamprianidis2018excitation}. CC BY 4.0.}
\label{Figure 8}
\end{figure}

Meanwhile, it is worth mentioning that, in all the above cases with plane wave illumination, the anapole states cannot make particles ideally 'radiationless' because of the reciprocity theorem. For a scattering object at anapole states, there are always some accompanied radiative terms $\it{Q}_{\ell m}$ or $\it{M}_{\ell m}$. Therefore, in most studies, the definition of 'nonradiating' anapole states is often relaxed to the local minima of $a_1$ in Mie theory or $E_{11}$ in charge-current expansion. Nevertheless, if a particle is excited by structured light with radial or azimuthal polarizations, other multipoles can be significantly or even entirely suppressed, thereby giving rise to a pure nonradiating anapole state. In Fig. \ref{Figure 8} we illustrate some interesting proposals to excite nearly perfect anapole states with structured light illumination \cite{raybould2017exciting, wei2016excitation, lamprianidis2018excitation}. For instance, while the anapole state in an isotropic dielectric nanosphere is generally not suitable to be observed under plane wave illumination \cite{miroshnichenko2015nonradiating}, it can be efficiently excited by Flying Doughnut pulses which can suppress other radiative higher-order multipoles \cite{raybould2017exciting} (see Fig. \ref{Figure 8}a). Another configuration was proposed by Wei \emph{et al.} \cite{wei2016excitation}, where they utilized two antiphased counter-propagating beams with radially polarizations and the same intensity, as shown in Fig. \ref{Figure 8}(b) and (c). In this way, they demonstrated that it is possible to externally excite an ideal anapole state without violating the reciprocity theorem, as the $4\pi$ illumination naturally fulfill the reciprocity condition $\int_{sph} \bf{E_f}^{4\pi} \cdot \bf{J} = 0$, where $\bf{E_f}^{4\pi}$ is the local field of the illumination in the absence of the particle and $\bf{J}$ is the induced polarization currents inside the particle. Thus, the scattered field outside the particle is hardly seen even in the near field. Very recently, Lamprianidis and Miroshnichenko systematically discussed scattering from different nanoparticles and proposed several schemes to excite anapole states of simple or hybrid nature \cite{lamprianidis2018excitation}.

\subsection{Enhanced near-field effects and optical nonlinearity}

Besides its significant far-field signature, anapole states also feature noticeable near-field enhancement and confinement of electric energy, as shown in Fig. \ref{Figure 3}, \ref{Figure 6}, and \ref{Figure 7}. It is worth noting that this phenomenon is usually achieved with lossless dielectrics rather than lossy metals. Therefore, in contrast to the conventional plasmonic 'hot spots' which are always associated with substantial Joule heating of nanostructures and local environment, the anapole-mediated hot spots in dielectrics provide their unique advantages for scenarios where localized heating is detrimental to applications or devices themselves. For instance, the heat generated by plasmonic nanostructures may vaporize the surrounding liquid or influence the properties of nearby emitters, molecules, or proteins \cite{mahmoudi2013variation, caldarola2015non}. The metallic nanoantennas would also be reshaped or even melted under strong illumination power, hindering their potential applications in several regimes such as nonlinear optics \cite{chen2012nanosecond}.  In addition, the 'transparent' characteristic induced by anapole states naturally offers new possibilities for noninvasive sensing or detection. 

\begin{figure}[t]
\centering
\includegraphics[width=15.5cm]{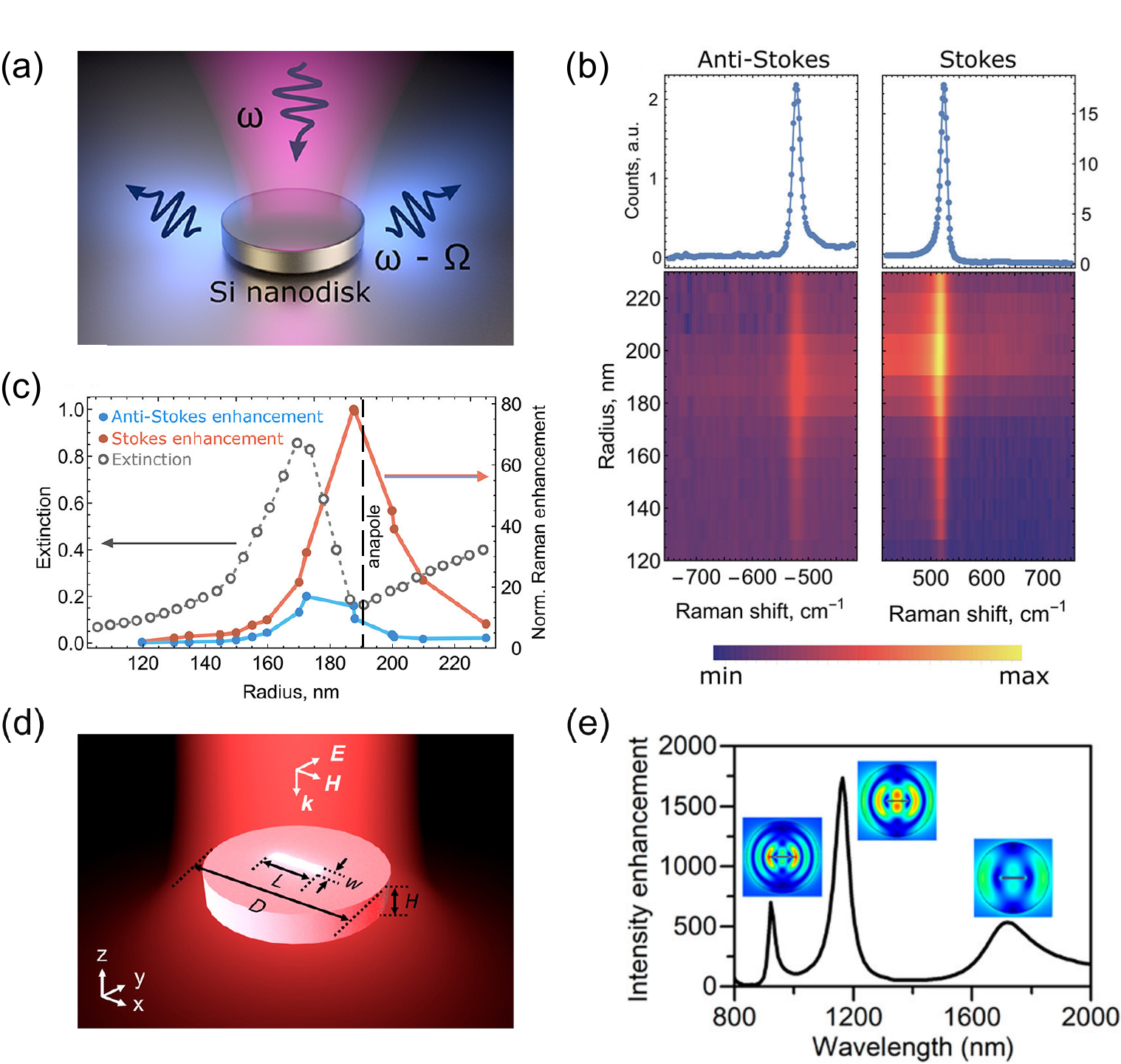}
\captionsetup{justification=justified, format = plain}
\caption{Anapole-assisted Raman scattering and near-field enhancements. (a) Schematic of anapole-enhanced Raman scattering in a silicon nanodisk. (b) Experimental Stokes and anti-Stokes Raman spectra from a 180 nm radius nanodisk (top) with 2D color maps of Raman spectra as a function of nanodisk radius for 785 nm excitation. (c) Extinction spectra (dashed line) measured at 785 nm for nanodisks with a varied diameter. A clear drop indicates the position of an anapole state. Solid lines represent integrated Stokes (red) and anti-Stokes (blue) Raman enhancement of the 522 cm$^{-1}$ phonon line under 785 nm excitation. (d, e) Illustration of anapole-induced near-field enhancement in a slotted silicon nanodisk and the multispectral intensity enhancement (e) of the system. (a-c) Reprinted with permission from \cite{baranov2018anapole}. Copyright (2018) American Chemical Society. (d, e) Reprinted with permission from  \cite{yang2018anapole}. Copyright (2018) American Chemical Society.}
\label{Figure 9}
\end{figure}

Fig. \ref{Figure 9}(a) shows the schematic of anapole-enhanced Raman scattering in silicon nanodisks, where anapole excitation and its enhanced light-matter interaction at the near field can amplify the inelastic scattering at Stokes-shifted frequency $\omega - \Omega$, with $\omega$ the incident frequency and $\Omega$ a phonon frequency. To verify this concept, Baranov \emph{et al.} \cite{baranov2018anapole} systematically measured the Raman response of silicon nanodisks with a fixed thickness of 70 nm and varying radii, as shown in Fig. \ref{Figure 9}(b). The strongest Raman enhancement at 785 nm excitation corresponds to the silicon nanodisk with a radius of 180 nm, which exactly agrees with the anapole position of the silicon nanodisks (Fig. \ref{Figure 9}c). Compared to an unstructured Si film, 2 orders of magnitude enhancement of Si Raman scattering intensity per volume unit can be achieved. The pronounced valley in the far-field scattering spectrum also indicates a suppression in elastic scattering. This unique response arising from the subradiant feature of anapole states is fairly distinct to conventional Raman scattering processes in nanostructures, where elastic and inelastic scattering are usually correlated and amplified simultaneously by radiant optical resonances \cite{stiles2008surface, cao2006enhanced, dmitriev2016resonant, alessandri2016enhanced}. 

In addition to Raman scattering, the enhanced near-field light-matter interaction at anapole states can be also used to amplify many other physical processes, such as realizing resonance strong coupling between anapole states and molecular excitons in J-aggregates \cite{liu2018resonance}. In these processes, the strengths of the effects are directly linked to the near-field enhancement factor achieved at anapole states. However, in conventional dielectric nanostructures, the field enhancements are inherently much lower than their plasmonic counterparts. The first-order anapole state in silicon nanodisks typically exhibits an intensity enhancement of electric fields around 5. Moreover, the strongest hot spot sitting in the center of the solid disks is not accessible by other physical objects such as nearby nanoemitters or biomolecules, which further limits the potential use of enhanced near fields at anapole states. To circumvent these issues, Yang \emph{et al.} \cite{yang2018anapole} designed a slotted silicon nanodisk to boost the near-field enhancement at anapole states by virtue of boundary conditions, while without bringing any degradations to their far-field signature. As shown in Fig. \ref{Figure 9}(c), when the incident light is polarized ($E_y$) perpendicular to the long axis of the slot (along the $x$ axis), the continuity of  the electric displacement $D_y$ enforces the electric field $E_y$ to abruptly increase along the slot-silicon interfaces and exhibit a dramatically higher amplitude inside the slot. In this way, the intensity enhancement of electric fields can easily exceed 1500 folds at anapole states with a multispectral response and an extra polarization dependence, which can be particularly favorable for nonlinear optics (Figure 9). The open slot region also provides exposed hotspots that are spatially overlapped with nearby quantum emitters, thereby leading to a radiative decay rate enhancement up to 800 folds with negligible Ohmic losses. The design principle of 'splitting field maxima' can be easily extended to other dielectric materials and structures. A very recent study \cite{mignuzzi2018nanoscale} demonstrated that by replacing the open slot with a delicate nanocavity, it is possible to further enhance the local density of optical states (LDOS) and radiative decay rate in GaP nanodisks at anapole states.

\begin{figure}[t]
\centering
\includegraphics[width=15.5cm]{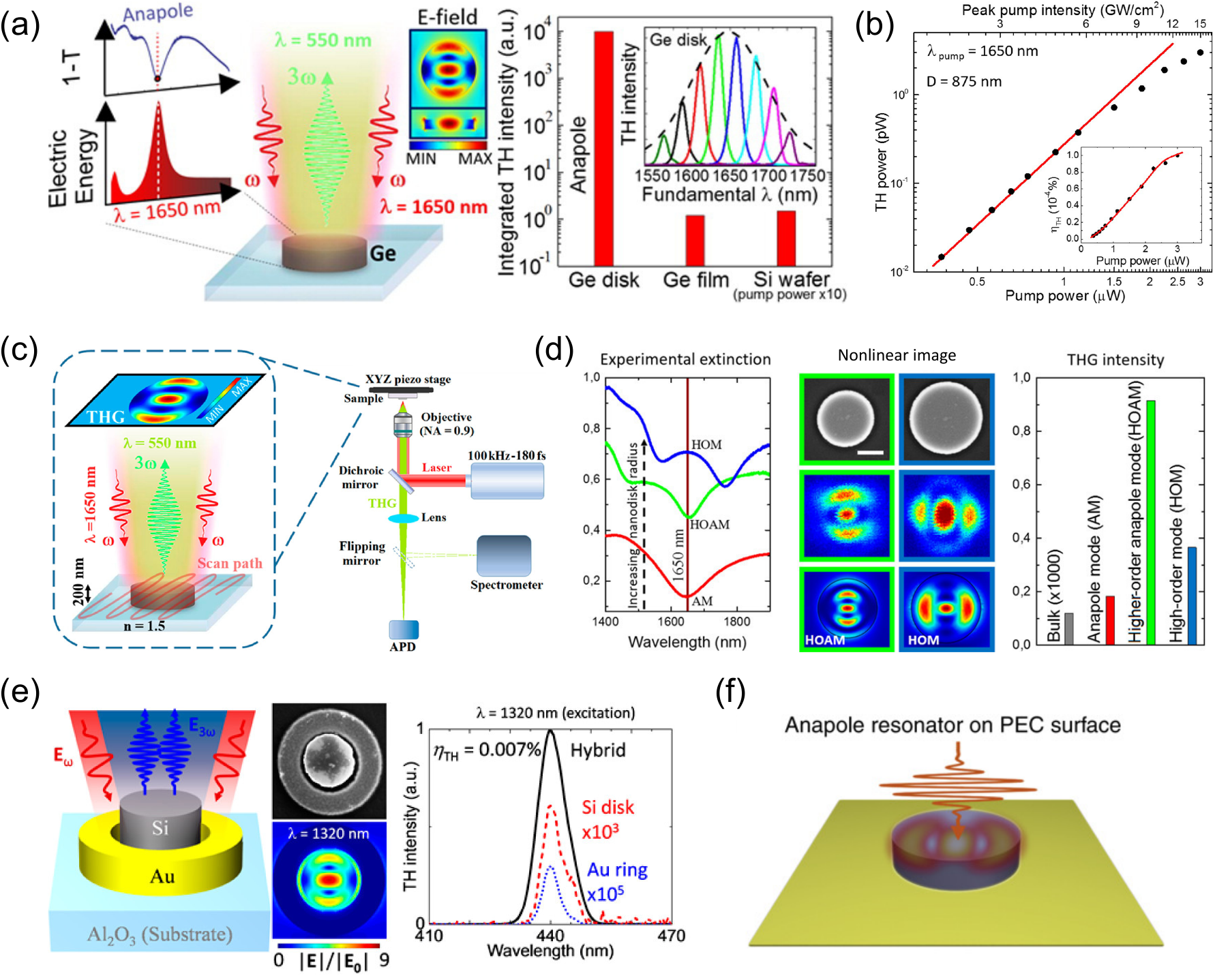}
\captionsetup{justification=justified, format = plain}
\caption{Enhanced harmonic generation by anapole states. (a) Third harmonic generation in individual Ge nanodisks excited at anapole states. (b) Experimentally measured third harmonic power as a function of pump power at anapole states. The insert shows the conversion efficiency versus pump power. (c) Schematic and optical setup for nonlinear imaging at higher-order anapole states. (d) Experimental extinction spectra, nonlinear images, and third harmonic generation at higher order anapole states in Ge disks. (e) Third harmonic generation in Si-Au hybrid nanodisks. (f) A configuration of a mirror-enhanced anapole resonator. (a, b) Reprinted with permission from \cite{grinblat2016enhanced}. Copyright (2016) American Chemical Society. (c, d) Reprinted with permission from  \cite{grinblat2016efficient}. Copyright (2016) American Chemical Society. (e) Reprinted with permission from  \cite{shibanuma2017efficient}. Copyright (2017) American Chemical Society. (f) Reprinted with permission from \cite{xu2018boosting}. CC BY 4.0.}
\label{Figure 10}
\end{figure}

As we have mentioned above, high-index dielectrics at anapole states can provide efficient enhancement of electric fields inside nanostructures while plasmonic materials solely offer field enhancement at metal surfaces. This marked difference endows dielectric nanoparticles with a better capability to couple light into inner volumes of the nanostructures and to utilize nonlinearities emerging from a material's bulk response. In this regard, germanium featuring a high refractive index ($>$ 4) and a large intrinsic third-order susceptibility in the near-infrared region \cite{zhang2014nonlinear} thus becomes a very promising material for nonlinear photonic devices. In 2016, Grinblat and Li \emph{et al.} demonstrated the first application of anapole states in nonlinear optics by utilizing Ge nanodisks and subsequently conducted several interesting works \cite{grinblat2016enhanced, grinblat2016efficient, shibanuma2017efficient}, as illustrated in Fig. \ref{Figure 10}. In their first experiment (Fig. \ref{Figure 10}a), they found that the third harmonic generation (THG) in Ge nanodisks can be greatly enhanced at anapole states due to the maximum electric energy inside the nanodisks. 4 orders of THG intensity enhancement was achieved compared to an unstructured Ge film. The power of the observed TH signal is proportional to the cube of the pump power with a maximum TH conversion efficiency up to 0.0001$\%$. We mention that this value achieved via anapole states was the highest efficiency of THG in individual nanoparticles at that time, exceeding 1 order of magnitude with respect to that of silicon disks driven at radiative magnetic resonances \cite{shcherbakov2014enhanced}. Soon later, they increased the TH conversion efficiency by another order of magnitude up to 0.001\% by utilizing higher-order anapole states \cite{grinblat2016efficient}.  By mapping the THG emission in the far-field, it is also possible to unveil the near-field intensity distribution of higher-order anapole states, as depicted in Fig. \ref{Figure 10} (c, d). This technique of nonlinear imaging offers a new possible way to examine other higher-order hybrid Mie resonances in dielectric nanostructures. 

The rapidly-showed encouraging prospect of anapole states in nonlinear optics soon attracted considerable attention \cite{zhai2016anticrossing, xu2018boosting, gili2018metal, timofeeva2018anapoles, rocco2018tuning}. More recent studies reported the great potential of hybrid nanostructures \cite{zhai2016anticrossing, xu2018boosting, gili2018metal}, i.e. incorporating dielectric nanoparticles at anapole states with auxiliary plasmonic nanostructures. A representative gold-silicon ring-disk structure is illustrated in Fig. \ref{Figure 10}(e), where the plasmonic resonance of the gold nanoring can boost the field enhancement at the anapole state of the silicon disk. A total TH conversion efficiency of 0.007\% was achieved at a wavelength of 440 nm, which can be tuned by modifying the geometrical parameters of the configuration \cite{shibanuma2017efficient}. Another approach to increasing the anapole response was to use a metal film as a mirror (see Fig. \ref{Figure 10}f). Different from most anapole states in dielectric nanoparticles which electric and toroidal dipole responses are usually far away from their resonances, the mirror-enhanced approach can generate anapole states consisting for resonant electric and toroidal dipoles due to coupling between displacement currents in dielectrics and the free electron oscillations within the metal surface. The significantly enhanced field inside the dielectric nanostructures makes it possible to reach the highest TH efficiency of 0.01\% to date \cite{xu2018boosting}, which is 2 orders of magnitude higher than that of an isolated dielectric disk on a glass substrate. Inspired by the advances in Ge or Si disks, researchers also explored the anapole states supported by nanostructures made of \MakeUppercase{\romannumeral 3}--\MakeUppercase{\romannumeral 5} materials such as GaAs, AlGaAs, or InAs, which have noncentrosymmetric structures and thus become an ideal platform for second harmonic generation (SHG) \cite{gili2018metal, timofeeva2018anapoles, rocco2018tuning}.  In particular, we mention the recent study by Timofeeva \emph{et al.} \cite{timofeeva2018anapoles}, in which they developed a novel FIB slicing technique to fabricate free-standing nanodisks and investigated their nonlinear response enhanced by anapole states. This special configuration is able to realize toroidal response and anapole states with different orientations and polarization selectivity. 

\begin{figure}[h]
\centering
\includegraphics[width=15.5cm]{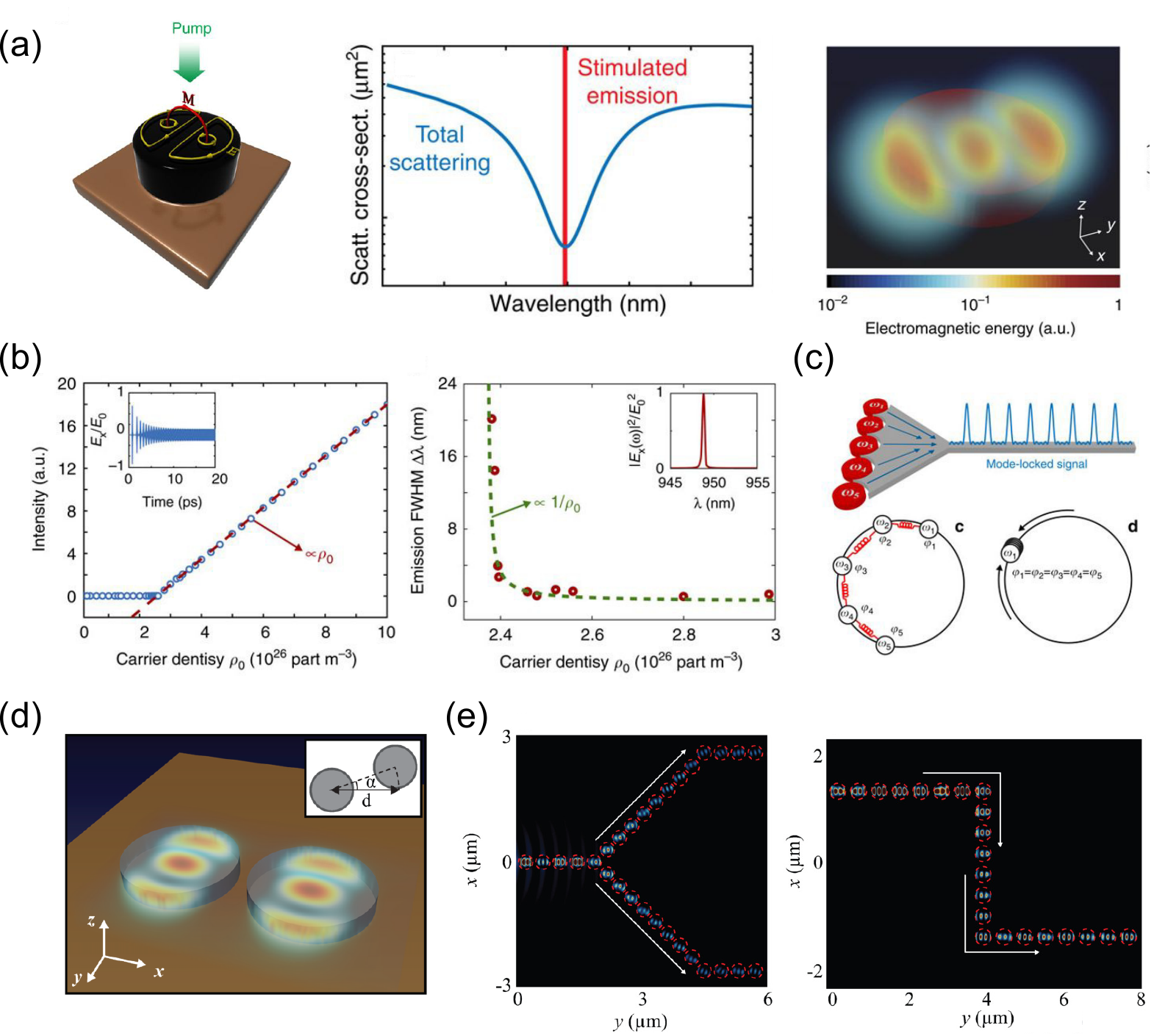}
\captionsetup{justification=justified, format = plain}
\caption{Near-field anapole nanolaser and energy guiding. (a) Concept of anapole laser composed of a direct gap semiconductor that is optically pumped (left), with its stimulated emission coinciding with the anapole state in the scattering spectrum (middle). 3D volume map of electromagnetic energy of a nanolaser at the steady state is also provided (right). (b) Intensity and linewidth of the electromagnetic field amplified inside the nanodisk with respect to carrier density $\rho_0$. The insets show time trace of field component $E_x$ at $\rho_0 = 3 \times 10^{26}\ \rm{part}\ m^{-3}$(left) and power spectral density inside the disk (right). (c) Configuration for mode-locking anapole laser with ultrafast pulse generation. The bottom two figures illustrate the working mechanism of the system, where nearest-neighbour coupling enables the anapole lasers operating at different emission frequency $\omega_n$ to mutually lock their phases into the same value. (d) Near-field mutual interaction of two silicon disks at anapole states. (e) Subwavelength guiding via near-field energy transfer of anapole states. (a-c) Reprinted with permission from \cite{gongora2017anapole}. CC BY 4.0. (d, e) Reprinted with permission from \cite{mazzone2017near}. CC BY 4.0.}
\label{Figure 11}
\end{figure}

Another important physical feature of many \MakeUppercase{\romannumeral 3}--\MakeUppercase{\romannumeral 5} compounds is their high carrier mobility and direct bandgaps, which allows them to serve as major constituents in active optoelectronic devices such as nanoscale lasers \cite{chen2011nanolasers, homewood2015optoelectronics, wan2016sub}. One interesting question naturally arises, i.e., whether it is possible to enhance stimulated light emission at anapole states. In a general sense, this idea is counterintuitive since the nonradiating nature of anapole states does not provide any conventional optical feedback that is visible in the far field. Nevertheless, this also heralds the emergence of a new type of near-field lasers that are radiationless in far field and free from the diffraction limit. The concept of an anapole laser \cite{gongora2017anapole} is given in Fig. \ref{Figure 11}(a). By adjusting the geometric parameters of a nanodisk composed of direct gap semiconductors, one can tune its anapole state coincide with the stimulated emission wavelength. In this way, a steady state can be generated with strongly confined energy that is only evanescently transmitted in a subwavelength region in the proximity of the nanodisk, which is radically different from a conventional classical laser. The amplitude and spectral linewidth of the anapole laser with respect to the pumping rate in an InGaAs disk are plotted in Fig. \ref{Figure 11}(b). The anapole intensity is linearly proportional to the carrier density $\rho_0$, following the typical behavior of a standard laser. The spectral width quickly reaches a stable value of $\sim$ 2 nm soon after the laser threshold. The insets also display the time evolution and frequency responses of the electric field. Furthermore, anapole mode locking and on-chip ultrafast pulse ($\sim$ 100 fs) generation can be achieved via near-field synchronization of different anapoles, as shown in Fig. \ref{Figure 11}(c). When a linear chain of anapole lasers emitting at slightly different frequencies $\omega_n$ and phases $\varphi_n$, the nearest-neighbor coupling of all these nonlinear oscillators would mutually lock their phases to the same value and generate a series of ultrafast pluses into the connected waveguide. The mutual coupling between anapole states in adjacent dielectric nanodisks also offers the possibilities to efficiently transfer and guide near-field energy with minimum scattering losses. An anapole-based waveguide \cite{mazzone2017near} have been demonstrated with superior robustness against physical bending and splitting compared to standard dielectric waveguides, as illustrated in Fig. \ref{Figure 11}(d, e). 

\subsection{Metamaterials and metasurfaces}

Metamaterials have spawned the first demonstration of dynamic anapoles in the electromagnetic spectrum. In turn, the emergence of anapole states set up a new physical playground for designing metamaterials and metasurfaces and extend their fascinating functionalities even further \cite{basharin2017extremely, wu2018optical, ospanova2018anapole, liu2017high, zhang2018high, han2018analog}. 

\begin{figure}[t!]
\centering
\includegraphics[width=14.5cm]{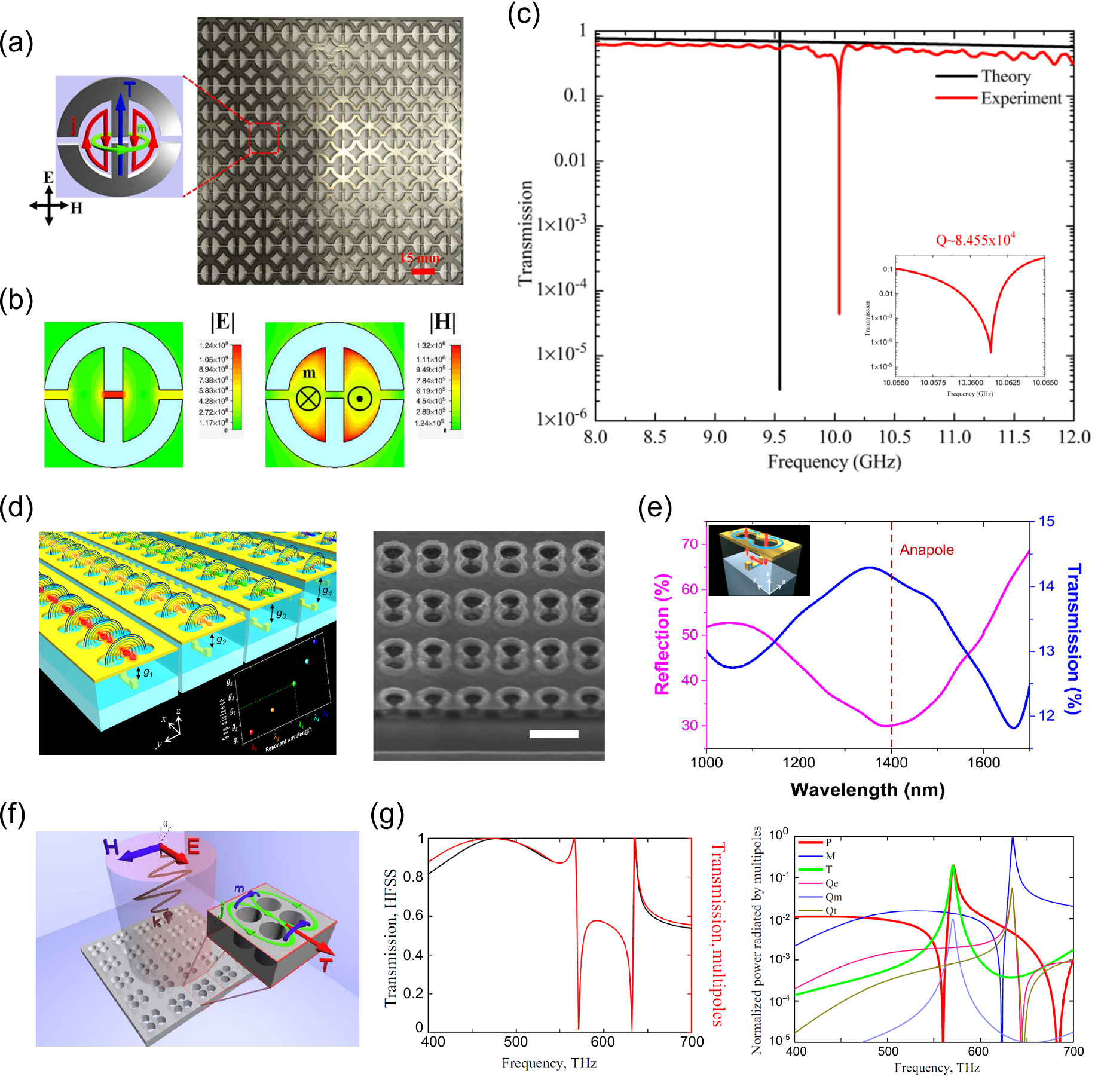}
\captionsetup{justification=justified, format = plain}
\caption{Anapole metamaterials. (a) Schematics of a microwave anapole metasurface and its constituent meta-atom. (b) Simulated electric and magnetic fields inside a meta-atom in the microwave anapole metamaterial. (c) Experimental and simulated transmission spectrum of the metamaterial with an extremely high Q-factor. (d) An optical anapole metamaterial consisting of plasmonic dumbbell apertures and vertical split-ring resonator underneath. (e) Reflection and Transmission spectra of the optical anapole metamaterial. (f) An optical anapole metamaterial with perforated holes on silicon. The formation mechanism of toroidal excitation is illustrated as $\bf{T}$ with $\bf{m}$ representing the magnetic dipole and $\bf{j}$ the induced electric displacement currents. (g) Transmission response and corresponding multipolar contribution in the far-field radiation of the silicon anapole metamaterials. (a-c) Reprinted with permission from \cite{basharin2017extremely}. Copyright (2017) American Physical Society. (d, e) Reprinted with permission from \cite{wu2018optical}. Copyright (2018) American Chemistry Society. (f, g) \cite{ospanova2018anapole} John Wiley \& Sons. Copyright (2018) WILEY-VCH Verlag GmbH \& Co. KGaA, Weinheim.}
\label{Figure 12}
\end{figure}

Several key studies are illustrated in Fig. \ref{Figure 12}. The first example in Fig. \ref{Figure 12}(a) shows a planar microwave metamaterial consisting of two symmetric split rings. When the incident electric field $\bf{E}$ is polarized along the central wire of the meta-atom, the excited current flow $\bf{j}$ will induce the circulating magnetic moments $\bf{m}$ around the central loops of the meta-atom, thereby giving rise to a strong toroidal dipole moment $\bf{T}$.  Meanwhile, a collocated electric dipole moment $\bf{P}$ will be excited by the central wire, making it possible to realize anapole excitation in the meta-atom and the composed metamaterials. This behavior is confirmed by numerical calculation of the electric and magnetic fields at anapole states of the meta-atoms, as shown in Fig. \ref{Figure 12}(b), where the electric field is strongly localized in the central gap while the magnetic field exhibits a closed torus distribution around the central wire. The transmission spectrum of the metamaterial shows an extremely high Q factor (over $8 \times 10^4$ in experiments and over $3 \times 10^6$ in simulations), which may find its use in modulators, sensors, and superconducting qubits.

Optical anapole metamaterials were also recently realized by Wu \emph{et al.} \cite{wu2018optical}, where they modified the structure of microwave dumbbell apertures into a planar and nanoscale version. The schematic of the configuration is shown in Fig. \ref{Figure 12}(d). Similar to the microwave dumbbell meta-atom, the dumbbell-perforated gold film can support a noticeable toroidal dipole. However, this toroidal response is not strong enough to dominate over other multipolar terms especially an accompanying magnetic quadrupole response. Therefore, an additional vertical split-ring resonator has to be implemented to break the symmetry, enhancing the toroidal response and suppressing the magnetic quadrupole moment. An observable anapole state thus could be achieved in such metamaterials with a clear peak in the transmission spectrum (and a clear dip in the reflection spectrum).

Similar mechanisms can be extended to dielectric metamaterials, as demonstrated by Ospanova \emph{et al.} \cite{ospanova2018anapole}. Fig. \ref{Figure 12}(f) shows a perforated silicon metamaterial which can be processed by one-step FIB fabrication. When the structure is illuminated from the top, the two loops of the induced displacement current $\bf{j}$ would create circulating magnetic dipoles $\bf{m}$ and thus a toroidal dipole $\bf{T}$. By interfering $\bf{T}$ and the electric dipole $\bf{P}$ generated by the holes themselves, an anapole state thus could be excited in the meta-atoms and the metamaterial. Two sharp peaks can be clearly observed in the transmission spectrum. Further multipole expansion unambiguously demonstrates that the two peaks correspond to the generalized Kerker condition \cite{liu2018generalized} ($\sim$ 630 THz) and an anapole generation ($\sim$ 570 THz), respectively. Thus, an optical metamaterial featuring multimode transparency can be obtained.  The hole configuration further makes it feasible to incorporate with any solvent media and function as a biosensor. 

\section{Conclusions and outlook}

Ever since its early theoretical predictions in atomic physics and electrodynamics, the nonradiating anapole state has puzzled scientists for decades. Recent studies have found its general existence in nanophotonics and subsequently unveiled its promising potential in many related areas such as cloaking, lasing, sensing, spectroscopies, metamaterials, and nonlinear optics. Here we provide a bird's-eye view on anapole states in nanophotonics research and hope that our timely review is able to show the vibrancy of the field. 

Meanwhile, despite the encouraging progress, the study of optical anapole states is still in its infancy, and we are witnessing new developments evolving at a rapid pace. So far, most of the studies have focused on individual nanoparticles with a disk geometry under ordinary planewave illumination. Additional degrees of freedom generating and controlling anapole states could be created by considering other structured geometries and/or employing structured light excitation (simultaneously). In this way, one may expect to realize anapole devices with much more flexible functionalities such as polarization dependence, chirality, and tailorable scattering directionalities. Another emerging opportunity lies in anapole metamaterials and metasurfaces \cite{krasnok2017nonlinear}, especially combining their proven advantages in nonlinear optics as well as the low Ohmic losses in dielectrics. For instance, a high-efficiency nonlinear metasurface realized via anapole states would be particularly unique and suitable for noninvasive sensing or detection, as its fundamental frequency naturally exhibits a ‘transparent window’ instead of a usual optical resonance.

Novel opportunities are also expected to appear with the extension of the current material palette. Further developments should also meet the needs and the overall trend of nanophotonics research, i.e. moving toward active and tunable devices with functionalities on demand \cite{zheludev2012metamaterials, koenderink2015nanophotonics}. An anapole metasurface with actively tunable responses was recently reported by using a structured phase-change alloy Ge$_2$Sb$_2$Te$_5$ \cite{tian2018dynamic}. Dynamic switching between nonradiative anapole states and radiative electric dipole modes was successfully demonstrated at different orders, rendering the metasurface a multispectral optical switch with a high contrast ratio over 6 dB. We expect that the future integration of anapole states into nanophotonic systems can shape a wide range of exciting applications, including optical sensing, nonlinear active metasurfaces, and many new types of near-field lasers, nanoantennas, and optical switches.

\section{Acknowledgements}
This work was funded by the European Research Council (the PLAQNAP project, Grant 341054) and the University of Southern Denmark (SDU2020 funding).

\bibliography{Ref_anapole}

\end{document}